\pdfoutput=1
\documentclass[pra]{revtex4-2}
\usepackage{float}
\usepackage{mathtools}
\usepackage{amsmath,bm}
\usepackage{amssymb}
\usepackage{amsmath} 
\usepackage{physics}
\usepackage{yfonts}
\usepackage{esint}
\usepackage{setspace}
\usepackage{subfigure}
\usepackage{soul}
\usepackage{dcolumn}
\usepackage{color}

\usepackage[hidelinks]{hyperref}

 \DeclareUnicodeCharacter{2212}{-}
\begin{document}

\title{Non-Hermitian elastic waveguides with piezoelectric feedback actuation: non-reciprocal bands and skin modes} 

\author{Danilo Braghini$^{a}$,  Luis G. G. Villani$^{b}$, Matheus I. N. Rosa$^{c}$, José R. de F. Arruda$^{a}$}
\affiliation{$^{a}$School of Mechanical Engineering, State University of Campinas, Campinas, São Paulo 13083-970, Brazil\\
$^{b}$Department of Mechanical Engineering, Federal University of Espírito Santo, Vitória, Espírito Santo 29075-910, Brazil \\ $^{c}$   Department of Mechanical Engineering, University of Colorado Boulder, Boulder CO 80309}

\date{\today}

\begin{abstract}
    In this work, we investigate non-Hermitian elastic waveguides with periodically applied proportional feedback efforts, implemented through piezoelectric sensors and actuators. Using one-dimensional spectral models for longitudinal motion, it is shown that dispersion diagrams of this family of structures exhibit non-reciprocal imaginary frequency components, manifesting as wave attenuation or amplification along opposite directions for all Pass bands. The effects of positive and negative proportional feedback, as well as local and non-local actuation are investigated. Overall, switching the sign of the feedback effort inverts the amplification direction, while increasing the degree of non-locality produces splitting of the Pass bands into multiple bands with interchanging non-reciprocal behavior. Furthermore, skin modes localized at the boundaries of finite domains are investigated and successfully predicted by the winding number of the complex dispersion bands. These results contribute to recent efforts in designing metamaterials with novel properties associated with the physics of non-Hermitian systems, which may find fruitful technological applications relying on vibration and noise control, wave localization, filtering and multiplexing. 
\end{abstract}

\maketitle


\vspace{2 cm}
This is the version of the article before peer review or editing. The published version can be found at \url{https://doi.org/10.1088/1361-6463/abf9d9}. To cite the version of record, use: : Danilo Braghini et al 2021 J. Phys. D: Appl. Phys. 54 285302

\newpage
\section{Introduction}
\label{sec: intro}

Phononics is a branch of condensed matter physics that has become a trending topic in recent years due to its applications in mechanical engineering, emanating from the design of phononic crystals (PC) and acoustic metamaterials~\cite{fok2008acoustic,hussein2014dynamics}. These artificially designed materials exhibit a series of intriguing properties, not commonly encountered in natural materials, which are promising for applications in vibration attenuation and noise reduction~\cite{huang2009wave,yang2010acoustic}, wave focusing~\cite{lin2009gradient}, cloaking,~\cite{cummer2007one}, and even as seismic barriers~\cite{miniaci2016large}. The discovery of topological insulators~\cite{hasan2010colloquium} has also inspired additional functionalities of metamaterials~\cite{huber2016topological,yang2015topological,ma2019topological}.  Those functionalities are particularly promising due to the robustness with respect to defects and disorder associated with the topological phenomena, with potential applications to robust signal processing~\cite{zangeneh2019topological} and multiplex communication~\cite{mei2019robust,ni2020robust}.
 
Recently, non-Hermitian (NH) systems have received considerable attention from the physics and engineering communities due to the realization of intriguing wave physics phenomena, such as single mode lasers~\cite{feng2014single}, unidirectional invisibility~\cite{lin2011unidirectional,fleury2015invisible}, Parity-Time (PT) symmetry and the properties of exceptional points~\cite{longhi2018parity,el2018non,miri2019exceptional}, and novel topological phases of matter~\cite{kunst2018biorthogonal,yao2018edge,shen2018topological,lee2019anatomy,ghatak2019new,torres2019perspective,kawabata2019symmetry,gong2018topological}. To an updated review, the reader is referred to \cite{bergholtz2019exceptional}. Particularly, NH edge states have already been realized on quantum-walk dynamics of single photons \cite{xiao2020non} and on electrical circuits \cite{helbig2020generalized}. A reciprocal version of this phenomenon was also observed \cite{hofmann2020reciprocal}. In the field of elastic/acoustic metamaterials, most investigations have focused on the properties of PT-symmetric systems~\cite{zhu2014p,christensen2016parity,liu2018unidirectional,hou2018tunable} where gain and loss are balanced, whose potential applications include asymmetric wave scattering~\cite{wu2019asymmetric}, higher-order topological insulators~\cite{zhang2019non,lopez2019multiple} and enhanced sensitivity of exceptional points~\cite{kononchuk2020orientation,rosa2021exceptional}. Outside of the scope of PT symmetry, recent works have aimed at exploring the properties of unbalanced systems with non-reciprocal interactions~\cite{brandenbourger2019non,rosa2020dynamics,ghatak2020observation,chen2020self,scheibner2020non,sirota2020non}. Among the key properties, large non-reciprocity associated with directional wave amplification and attenuation has been demonstrated, along with novel topological modes caused by the NH skin effect \cite{yao2018edge,lee2019anatomy,helbig2020generalized}, whereby a large number of the bulk modes of a 1D system become localized at a boundary. These works illustrate a variety of properties associated with NH physics, which are not found in Hermitian (conservative) counterparts and, hence, may pave the way to the design of active metamaterials with novel functionalities.

In this context, the use of feedback interactions has been a common theme in the study of active NH mechanical systems~\cite{brandenbourger2019non,rosa2020dynamics,chen2020self,sirota2020non}. Although already proposed by a few recent works, the study of the wave dynamics of elastic media with feedback interactions is still in its earlier stages, and much more work is needed to understand their fundamental properties and practical applicability. Towards filling this gap, this paper proposes the study of non-Hermitian elastic waveguides with periodically applied proportional feedback efforts, implemented through piezoelectric sensors and actuators. Piezoelectric materials such as PZT have been widely explored for vibration control and  energy harvesting~\cite{sodano2004review,anton2007review,bhalla2016piezoelectric}, in phononics~\cite{thorp2001attenuation,hou2004phononic,benchabane2006evidence,achaoui2011experimental,aly2018significance}, and for more complex applications such as non-reciprocal wave motion in spatio-temporal modulated waveguides~\cite{marconi2020experimental} and topological pumping of edge states~\cite{xia2020experimental}. They have also been employed in the exploration of NH systems, such as in PT symmetric waveguides~\cite{wu2019asymmetric,hou2018tunable}, exceptional points with enhanced sensitivity~\cite{kononchuk2020orientation,rosa2021exceptional} and metamaterials with odd micropolarity~\cite{chen2020self}. As such, PZTs are excellent candidates for the design of active metamaterials, although their use to establish non-reciprocal feedback interactions in NH systems remain limited. 

Herein, we consider a periodic NH one-dimensional elastic waveguide undergoing longitudinal motion, whereby PZT rod elements are interchangeably used as sensors and actuators, allowing for several feedback strategies to be implemented. In particular, we consider proportional feedback schemes where the voltage applied to each actuator is proportional to the voltage read by an adjacent sensor or by a sensor a few units away, which are respectively labeled as local and non-local feedback schemes. Aligned with recent findings in discrete lattice systems~\cite{rosa2020dynamics}, we observe that the family of NH waveguides exhibits largely non-reciprocal behavior, manifesting as wave amplification and attenuation along opposite directions for all the  Pass bands. We also illustrate that the non-locality of the feedback schemes is associated with a splitting of the Pass bands into multiple bands with opposite non-reciprocal behavior. Finally, NH skin modes localized at the boundaries of finite waveguides are also demonstrated. Their existence is successfully predicted by the winding number of the dispersion bands, and further confirmed by forced response simulations. 

\section{Non-Hermitian piezoelectric elastic rods}
\begin{figure}[H]
\centering
{\includegraphics[width =0.6\textwidth]{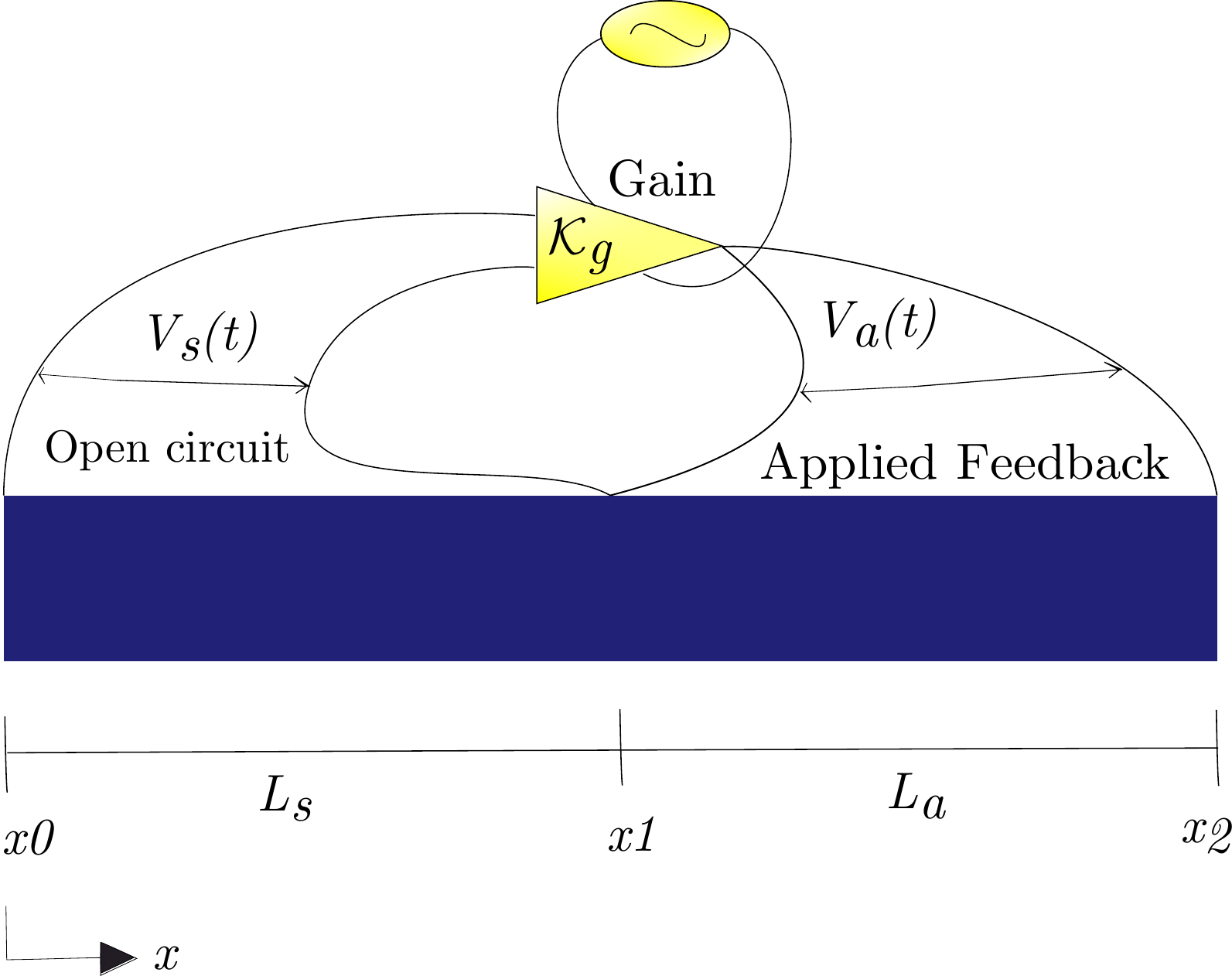}}
\caption{ \textbf{Unit cell made of local feedback-based phononic rod}. An amplifier
generates a voltage at the terminals of the actuator ($L_a$), which is proportional to the voltage at the terminals of the sensor ($L_s$). }
\label{modelcont}
\end{figure}

\raggedbottom
We consider a one-dimensional piezoelectric rod. Its unit cell, displayed in Fig.~\ref{modelcont}, comprises two segments with coupled mechanical and electrical degrees of freedom. Considering uniform cross section and material properties, the homogeneous equation of motion of the piezoelectric rod can be expressed as \cite{wang2008wave}
\begin{equation}\label{eq: waveeq_1}
  Y \pdv[2]{u (x,t)}{x} + e \pdv[2]{\varphi(x,t)}{x} - \rho \pdv[2]{ u(x,t)} {t} = 0,
\end{equation}
\begin{equation}\label{eq: waveeq_2}
  e \pdv[2]{u (x,t)}{x} - \alpha \pdv[2]{\varphi(x,t)}{x} = 0,
\end{equation}
where $u(x,t)$ is the axial displacement, $\varphi(x,t)$ is the electric potential, $Y$ is the Young's modulus, $\rho$ represents mass density, $\alpha$ is the dielectric constant, $e$ is the piezoelectric constant, $x$ is the axial coordinate and $t$ represents time. Additionally, one can derive the constitutive equations related to the mechanical and electrical internal efforts, respectively \cite{li2016analysis}
\begin{equation}\label{internal_N}
   N (x,t) = Y A\pdv{u(x,t)}{x} + e A \pdv{\varphi (x,t)}{x},
\end{equation}
\begin{equation}\label{internal_Q}
    Q (x, t) = e A \pdv{u(x,t)}{x} - \alpha A \pdv{\varphi (x,t)}{x},
\end{equation}
where $N(x,t)$ is the axial internal effort, $Q (x,t)$ is the electric charge distribution and $A$ represents the cross-section area. These equations completely define the wave propagation behavior in the piezoelectric rod domain, and can be solved using different numerical methods. In this work, we consider the Finite Element Method (FEM) and the Spectral Element Method (SEM), with formulations derived in appendix.

To realize a feedback-based unit cell, the direct piezoelectric effect is considered for a sensor with open circuit electric boundary condition, while the inverse effect is used to realize the actuator, by making the voltage applied to the actuator ($V_a (t)$) proportional to the voltage measured at the sensor boundaries ($V_s (t)$). Proportional feedback is represented by the equation:
\begin{equation}
    \kappa_g = \frac{V_a(t)}{V_s(t)} = \frac{K_g \alpha}{e},
    \label{eq: controllaw}
\end{equation}
where $\kappa_g$ is defined as the voltage gain, while $K_g$ is the gain in terms of the relation between $V_a$ and the difference in the sensor end displacements (see equation (\ref{Kg})). 

The voltage $V_s(t)$ in the sensor can be obtained by integrating   (\ref{internal_Q}) (\cite{li2016analysis}) over the sensor domain  ($Q(x,t) = 0$) :
\raggedbottom
\begin{equation}
     V_s(t) =  \int_{0}^{L_s} \pdv{\varphi (x,t)}{x} dx =  \frac{e}{\alpha} \left[u(L_s,t) -u(0,t) \right],
     \label{voltage_sensor}
\end{equation}
which depends on the measured displacements at the ends of the piezoelectric sensor. 

\begin{figure}[H]
\centering
{\includegraphics[width=\textwidth]{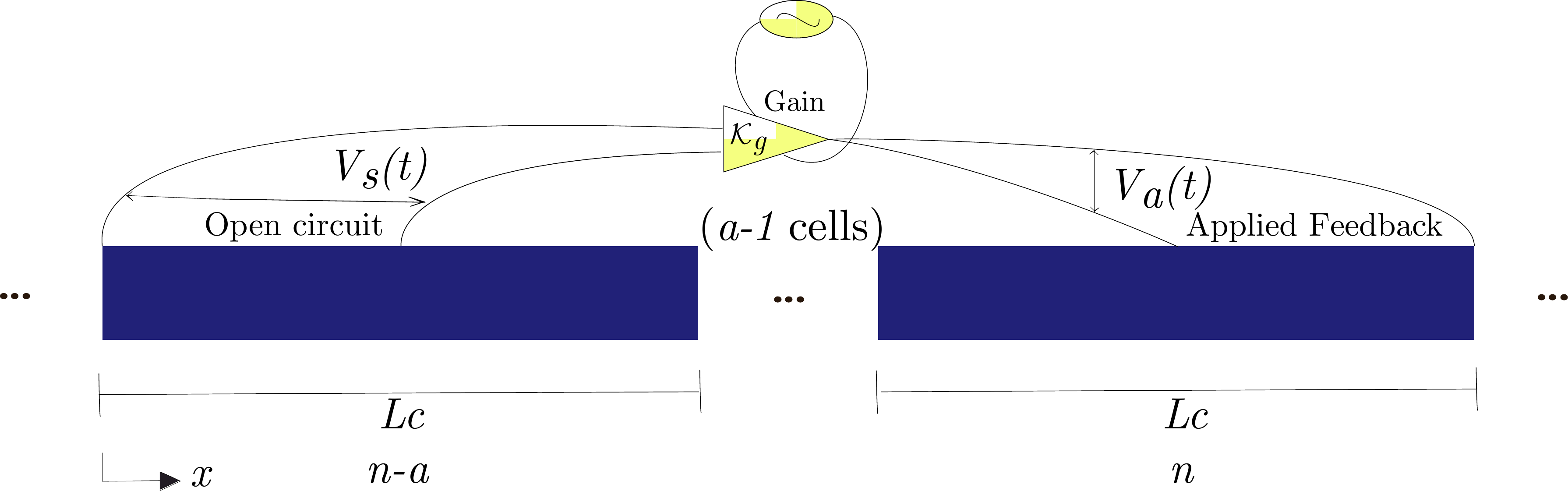}}
\caption{ \textbf{Non-local feedback scheme for a phononic crystal with $a > 0$}. The figure only shows the circuits of the actuator of the $n$-th cell and the sensor of the previous cell, $n-a$, for they have a feedback interaction.}
\label{nonlocal}
\end{figure}

In addition to the gain  $\kappa_g$, whose sign defines if the feedback interaction is negative or positive, we are also interested in the influence of the locality parameter $a$. As illustrated in Fig.~\ref{nonlocal}, the voltage applied to an actuator is proportional to the voltage measured by a sensor located $a$ units behind it. The schematic in Fig.~\ref{modelcont} illustrates the local feedback case, for which $a=0$. The general expression for the feedback interaction relating the voltage at the actuator and displacements at the sensor in the frequency domain is given by:

\begin{equation*}
    \hat{V}_a^n
    =
    K_g
    \begin{bmatrix}
    -1 &  1
    \end{bmatrix}
    \begin{bmatrix}
    \hat{u}(0)\\\hat{u}(L)
    \end{bmatrix}_s^{n-a},
\end{equation*}
which, by the \textit{Bloch-Floquet} theorem, becomes the following relation in terms of the $n$-th cell variables only

\begin{equation*}
    \hat{V}_a^n
    =
    K_g e^{i a k L_c}
    \begin{bmatrix}
    -1 &  1
    \end{bmatrix}
    \begin{bmatrix}
    \hat{u}(0)\\\hat{u}(L)
    \end{bmatrix}_s^n.
\end{equation*}

\raggedbottom

A finite structure can then be constructed by assembling a number of these cells. The dynamics of the structure can be predicted by the wave propagation features of the corresponding periodic (infinite) material \cite{hussein2006dispersive}, unveiled by the dispersion relation. We calculate the propagating wave modes of the dispersion relation $\omega (k)$, where $\omega$ is the angular frequency and $k$ is the wavenumber, by imposing real values of the wavenumber within a \textit{Brillouin} zone (BZ) and computing the corresponding $\omega$ values. On the other hand, to obtain the normal modes of the finite structure with a given number of cells we used the FEM method to assemble global stiffness and mass matrices and then solved the generalized eigenvalue problem related to free vibration. We choose to show the results in terms of $f =\frac{\omega}{2 \pi}$. The FEM model was also used to compute transient responses %to assigned forcing 
through standard numerical integration and to validate the implemented SEM solution (see appendix). Finally, Tab. \ref{tab:par} presents all the parameters used in the simulations.% performed.

Dissipation was included by assuming Rayleigh damping, i.e., assuming that the damping matrix is proportional to the stiffness and mass matrices. Using a reference damping 
ratio ($\zeta$) equal for all frequency ranges considered, it is possible to calculate the approximate constants of proportionality by applying the strategy described by Hall \cite{hall2006problems}. This strategy facilitates the comparison between the results obtained using the FEM and SEM models. In the latter, damping is added as an imaginary part of the Young's modulus proportional to the loss factor $\eta$. In addition to improving numerical conditioning, the use of dissipation is also very common in structural dynamics, since models become asymptotically stable and more representative of an experimental scenario. 

 \begin{table}[H]
	\centering
	\caption{Parameters used in the simulations (PZT - 5H).}
	\begin{tabular}{c c}
		\textbf{Parameter} & \textbf{Value} \\
		\hline
		Young’s modulus ($Y$) & 117 [GPa] \\
	    Piezoelectric constant ($e$) & 23.3 [C/m$^2$] \\
	    Density ($\rho$) & 7500 [kg/m$^3$] \\
	    Dielectric constant ($\alpha$) & 13.02 $\times 10^{-9}$ [F/m] \\
	    Cross section area ($A$) & 0.0028 [m$^2$] \\
	    Sensor length ($L_s$) & 0.005 [m] \\
	    Actuator length ($L_a$) & 0.005 [m] \\
	    %Damping 
	    Loss factor ($\eta$) & 0.001  \\ %\%
	    Number of cells of the finite structure & 40 \\
	    Boundary condition of the finite structure & free-free \\
	    \hline
    	\end{tabular}
\label{tab:par}
\end{table}

\section{Local feedback}
\label{results-local}

We begin by investigating the behavior of NH waveguides with local feedback ($a=0$), while the features associated with non-local feedback ($a>0$) are presented in the next section. The dispersion relations are first analyzed, followed by transient simulations confirming the predicted non-reciprocal behavior. Next, finite structures are investigated whereby the NH skin effect is observed.

\subsection{Non-reciprocal wave propagation}

Wave propagation in the NH waveguides is predicted by computing the dispersion relations by two different methods. One is a semi-analytical method based on the transfer matrix (TM) via the spectral element formulation (SEM) \cite{doyle1989wave}, which relies on exact solutions in the frequency domain. The other is a finite element approximation (FEM), which uses a finite number of degrees of freedom (DoF) through polynomial shape functions that approximate the displacement field \cite{craig2006fundamentals}. A numerical analysis comparing the methods and validating the mathematical models can be found in appendix. 

Figure (\ref{unfoldedDD}) presents the dispersion diagrams for a representative case with $\kappa_g= - 2$. The dashed lines correspond to the multiple solutions obtained as a function of the wavenumber $k$, which exhibits multiple folds, as expected from the application of Bloch's theorem. The solid black lines correspond to the correct dispersion branches selected in each BZ (the $n-$th band is selected in the $n$-th BZ). As further illustrated in Appendix, this procedure avoids the misinterpretation of both the imaginary frequency component, and the propagating wavenumber that could be folded into another BZ.

\begin{figure}[H]
\centering
\subfigure[]{\includegraphics[width=0.495\textwidth]{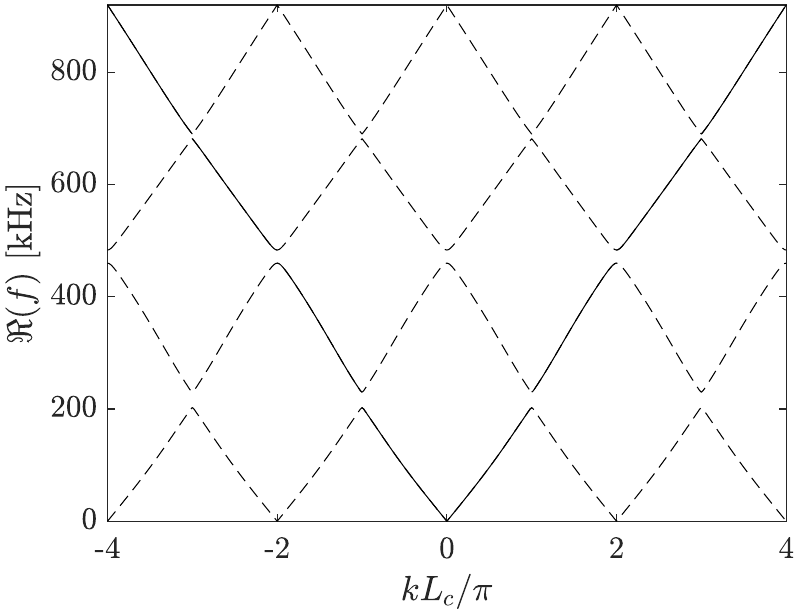}\label{2a}}
\subfigure[]{\includegraphics[width =0.495\textwidth]{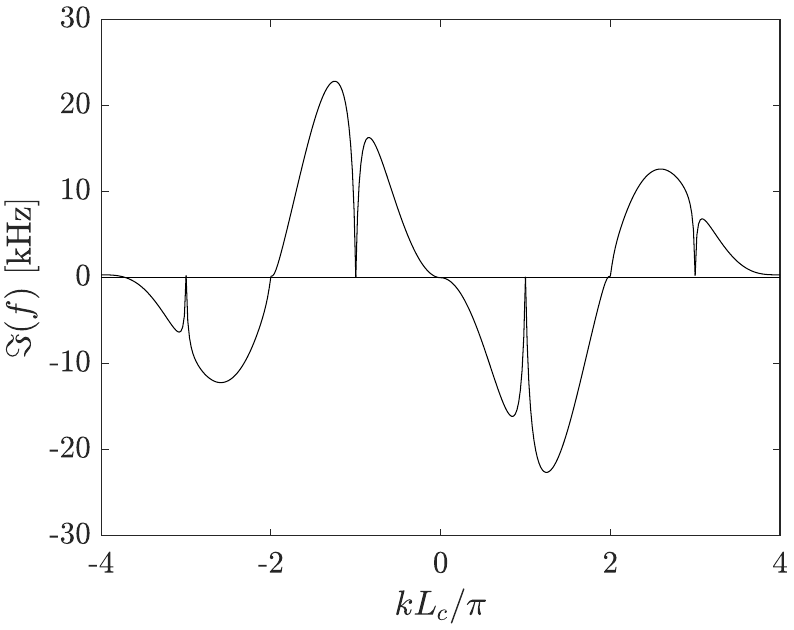}\label{2b}}
\caption{\textbf{Unfolded dispersion diagrams from the cell of Fig.~\ref{modelcont}}. (a) Real part of $f$ versus normalized wavenumber. (b) Imaginary part of $f$ versus normalized wavenumber.}
\label{unfoldedDD}
\end{figure}

\begin{figure}[H]
\centering
\subfigure[]{\includegraphics[width=0.495\textwidth]{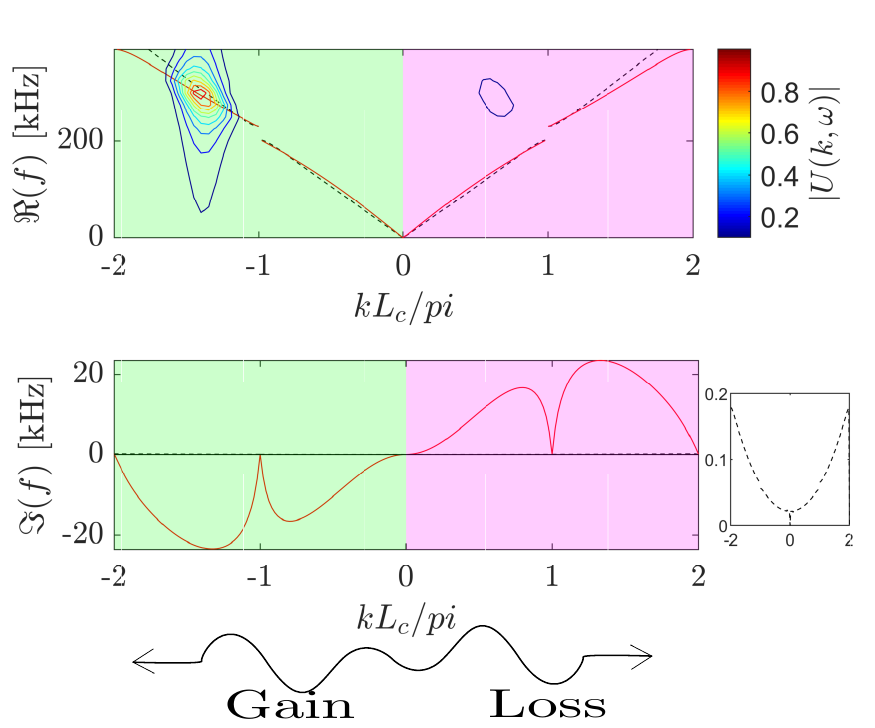}\label{4a}}
\subfigure[]{\includegraphics[width =0.495\textwidth]{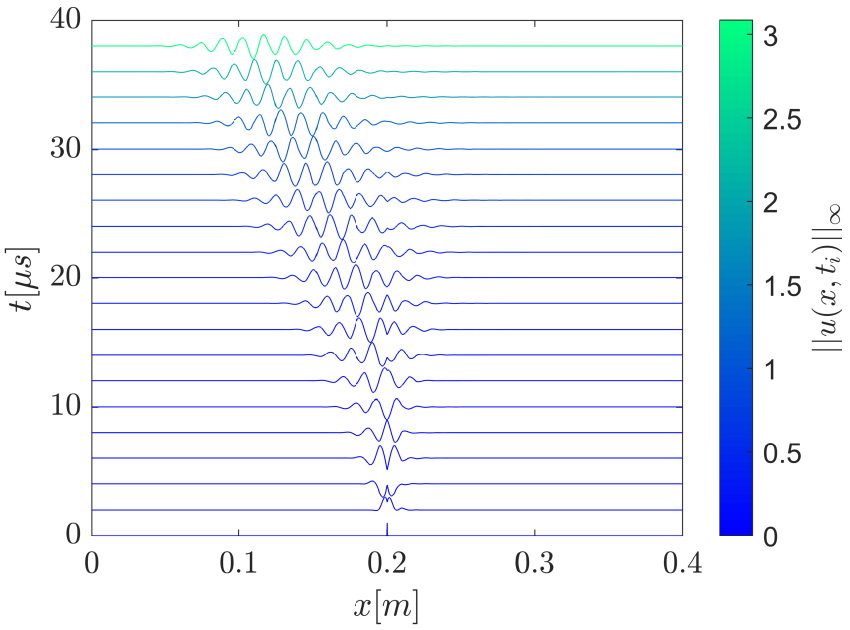}\label{4b}}
\subfigure[]{\includegraphics[width =0.495\textwidth]{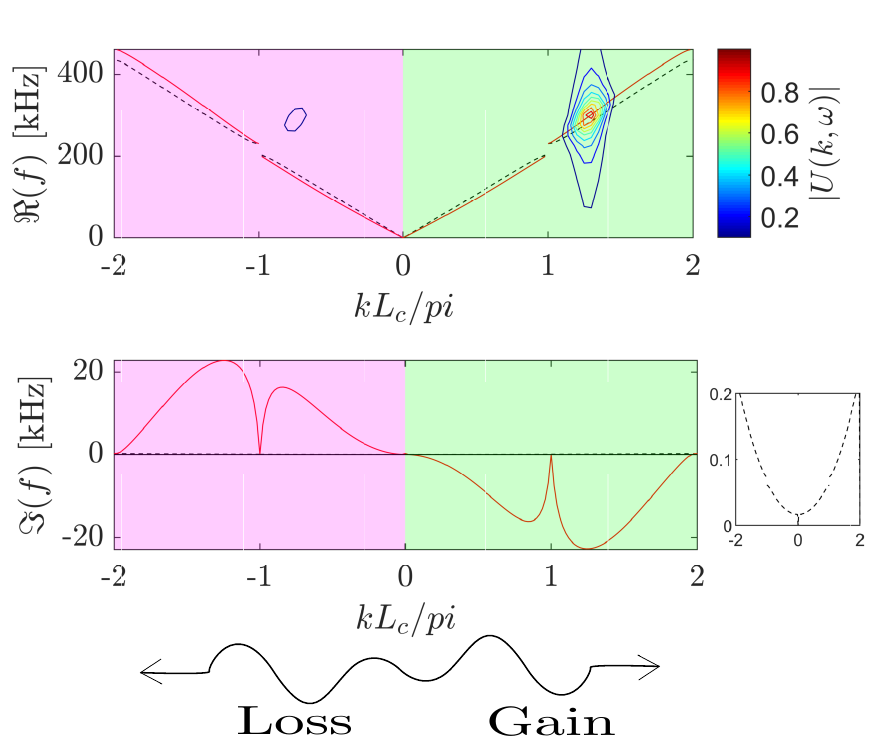}\label{4c}}
\subfigure[]{\includegraphics[width =0.495\textwidth]{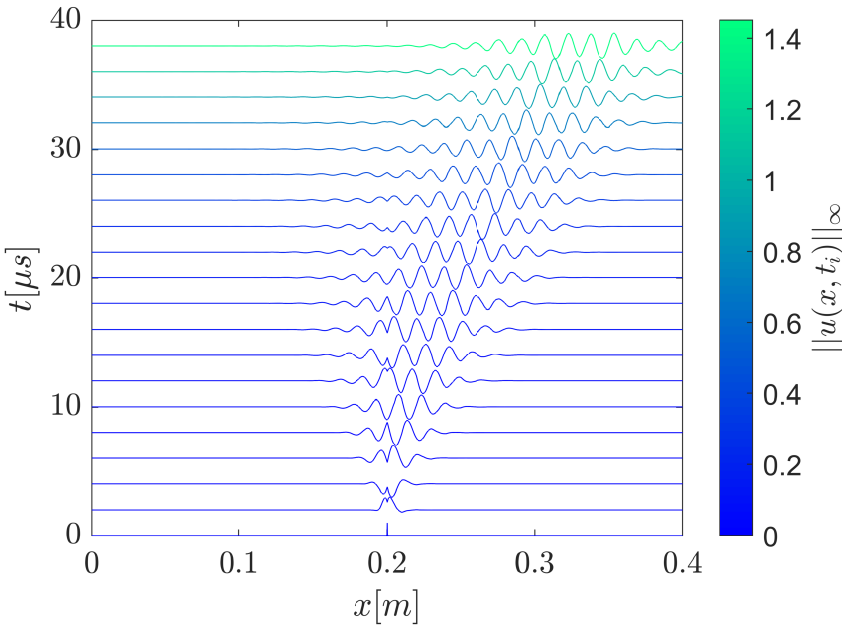}\label{4d}}
\caption{\textbf{Non-reciprocal amplification and attenuation of waves in structures with local feedback interactions ($a = 0$)}. (a) and (c) Propagation modes of the dispersion relation for $\kappa_g = 1.5$ and $\kappa_g = -2$, respectively. First and second PBs are shown in solid red lines, superimposed to results for the equivalent passive lattice (making $\kappa_g = 0$) displayed in dashed black lines. Attenuation and amplification zones are identified, respectively, by shaded
magenta and green areas.The inlets show zoomed-in plots of the imaginary dispersion diagrams to allow visualizing the small amplitude reciprocal results for damped passive lattices. (b) and (d) Transient responses to sine-burst with center frequency $300$kHz illustrating the non-reciprocal wave propagation for $\kappa_g = 1.5$ and $\kappa_g = -2$, respectively. Non-reciprocal wave propagation is further confirmed by their dispersion estimation through 2DFFT displayed as contours in (a) and (c), which are normalized by their maximum value. %(e) and (f) Zoom on the imaginary dispersion diagrams to show the small and reciprocal results for damped passive lattices
.}
\label{fig:4}
\end{figure}

The dispersion diagram is interpreted based on the general wave propagation solution $u(x,t)=U(k,\omega) \; exp[i(\omega t-k x)]$. By the convention adopted, $k > 0$ implies waves propagating to the right, whereas $k < 0$ indicates propagation to the left. The imaginary frequency component is related to exponential attenuation or amplification in time: $\operatorname{\mathfrak{I}}(f) > 0 $ is associated with attenuation, while $\operatorname{\mathfrak{I}}(f)  < 0 $ to amplification. With this in mind, Fig.~\ref{2b} unveils that each Bloch band (PB) amplifies waves propagating in one direction, while attenuates the waves that propagate in the opposite direction. This is further illustrated in Fig.(\ref{4a}) and Fig.(\ref{4c}), which present the dispersion for $k_g=1.5$ and $k_g=-2$ (solid red lines),  respectively, where shaded magenta and green areas indicate attenuation and amplification bands. We note that switching the sign of the feedback gain inverts the direction of amplification. Furthermore, dashed black lines are used to represent the dispersion of the equivalent passive lattice ($k_g=0$) for comparison. The presence of dissipation makes the passive lattice non-Hermitian with a small imaginary frequency component (see insets on  Fig.(\ref{4a}) and Fig.(\ref{4c})). Nevertheless, this component is entirely reciprocal- as can be seen by the symmetry with respect to the wavenumber- while the active systems with feedback exhibit entirely non-reciprocal attenuation and amplification features. 

The non-reciprocal behavior is confirmed by observing the transient response of finite domains. We consider a structure made of $40$ unit cells, with an excitation force applied at the middle of the structure. The excitation is a band-limited sine burst signal with 15 cycles and center frequency $f_c = 300$ [kHz], which excites primarily the second PB. Transient responses shown in Fig.~\ref{4b} and Fig.~\ref{4d} for positive and negative local feedback cases, respectively, confirm the previous analysis and illustrates the non reciprocity of this type of system. In particular, the targeted second PB amplifies waves traveling to the left for positive feedback, and to the right for negative feedback. We note that the waterfall plots are normalized by the maximum displacement ($\mathcal{L}_{\infty}$ norm) along the rod for each time instant $t_i$ %and scaled by a factor of $10^9$ 
to improve visualization, while the corresponding color maps display the instantaneous $\mathcal{L}_\infty$ norm multiplied by a factor of $10^9$ to illustrate the amplification. On the dispersion diagrams of Fig.~\ref{4a} and Fig.~\ref{4c}, we have also superposed the two-dimensional discrete Fourier transform (2DFFT) of the transient response signal, showing the agreement with the predicted dispersion curves.

\subsection{Skin modes and localized vibration}
\begin{figure}[H]
\centering
\subfigure[]{\includegraphics[width=0.495\textwidth,height=0.25\textheight]{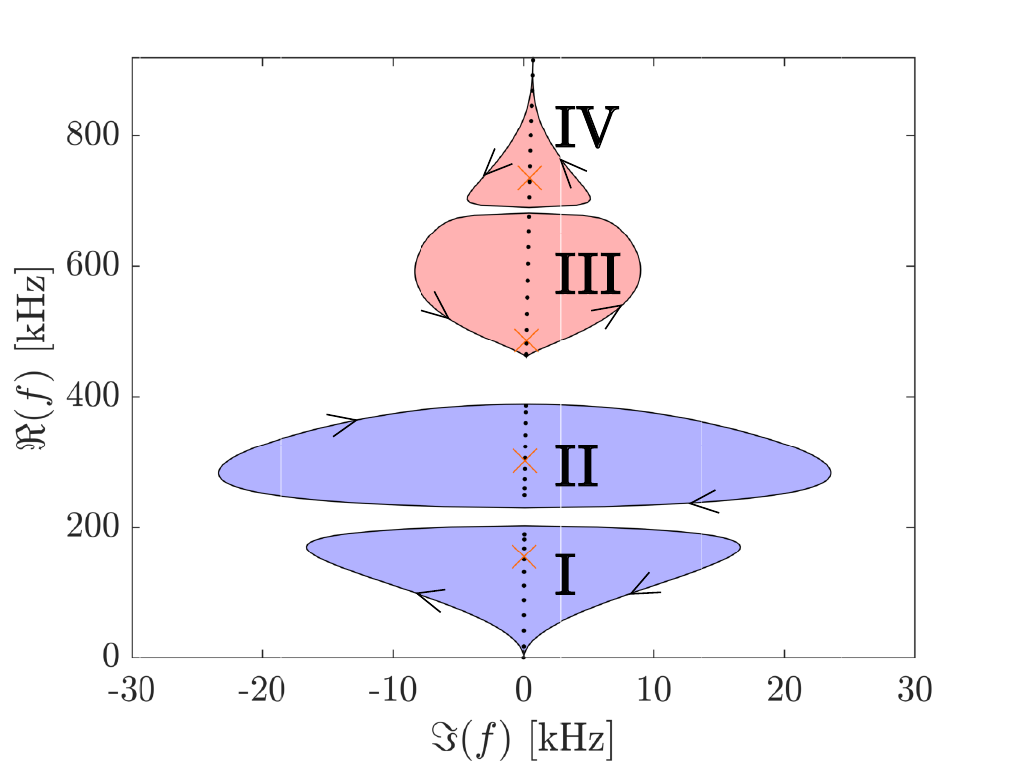}\label{6a}}
\subfigure[]{\includegraphics[width=0.495\textwidth,height=0.25\textheight]{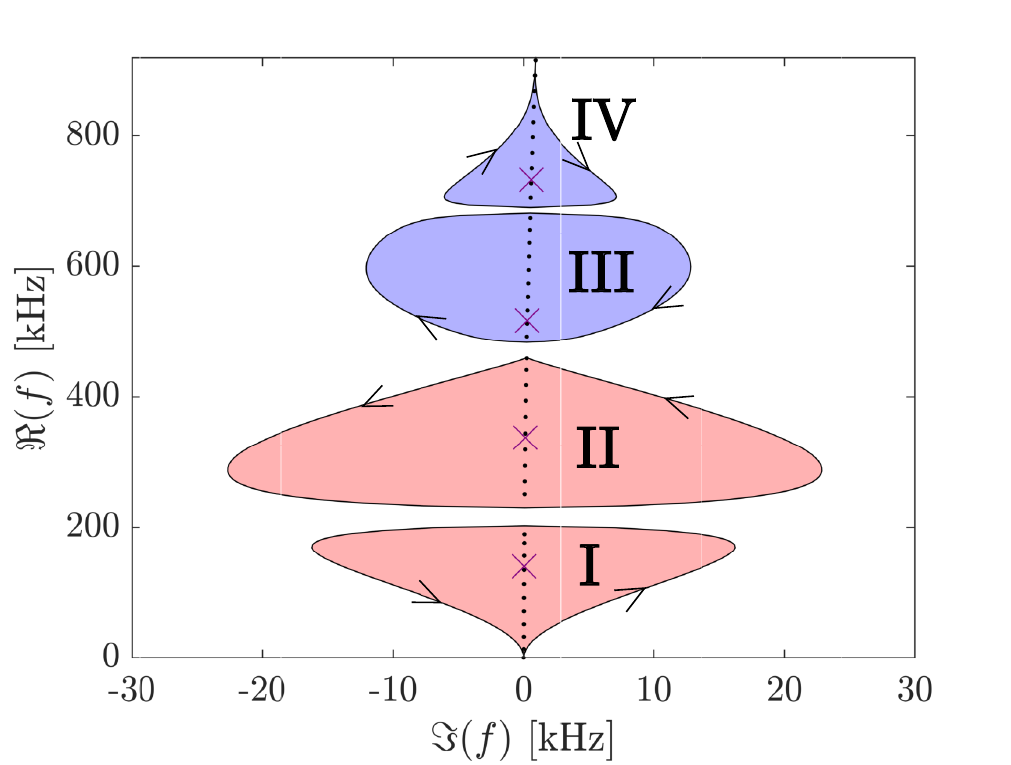}\label{6b}}
\subfigure[]{\includegraphics[width=0.495\textwidth,height=0.25\textheight]{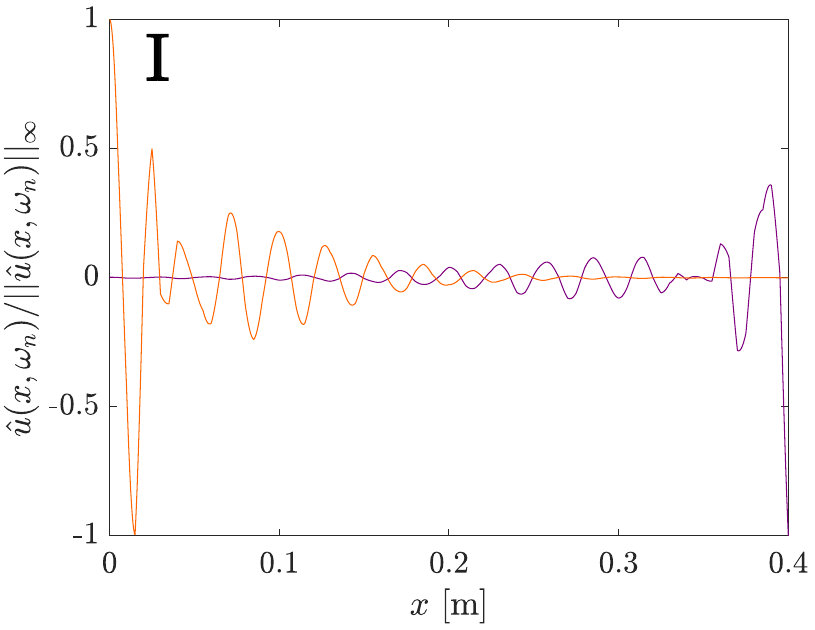}\label{6c}}
\subfigure[]{\includegraphics[width=0.495\textwidth,height=0.25\textheight]{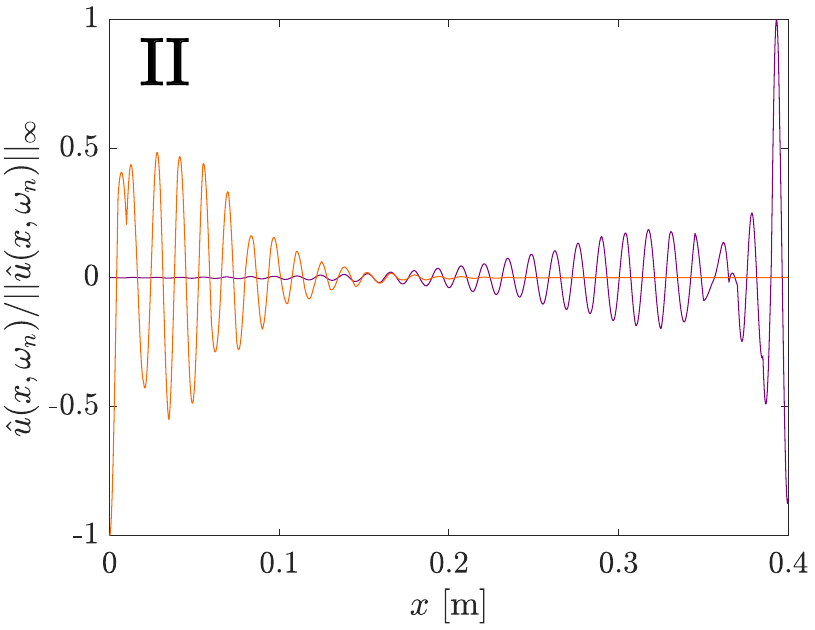}\label{6d}}
\subfigure[]{\includegraphics[width=0.495\textwidth,height=0.25\textheight]{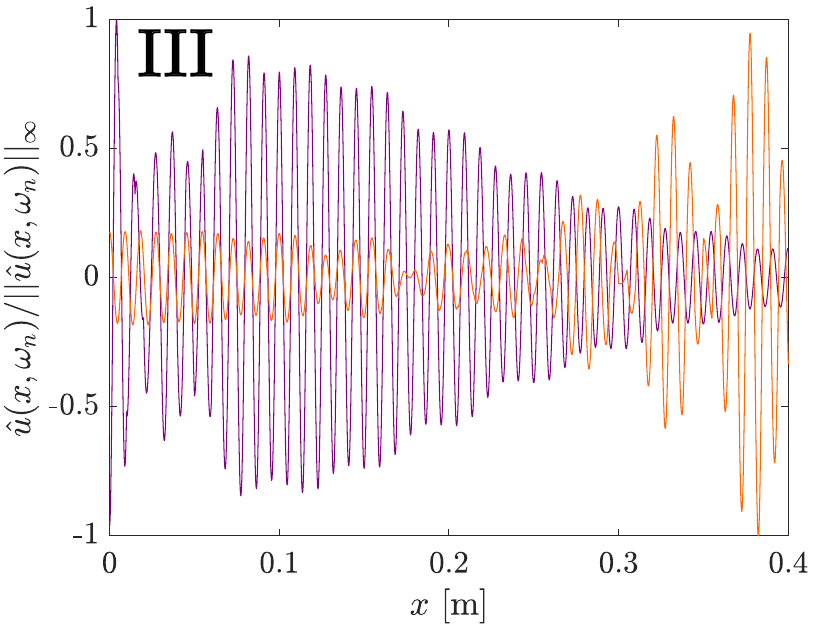}\label{6e}}
\subfigure[]{\includegraphics[width=0.495\textwidth,height=0.25\textheight]{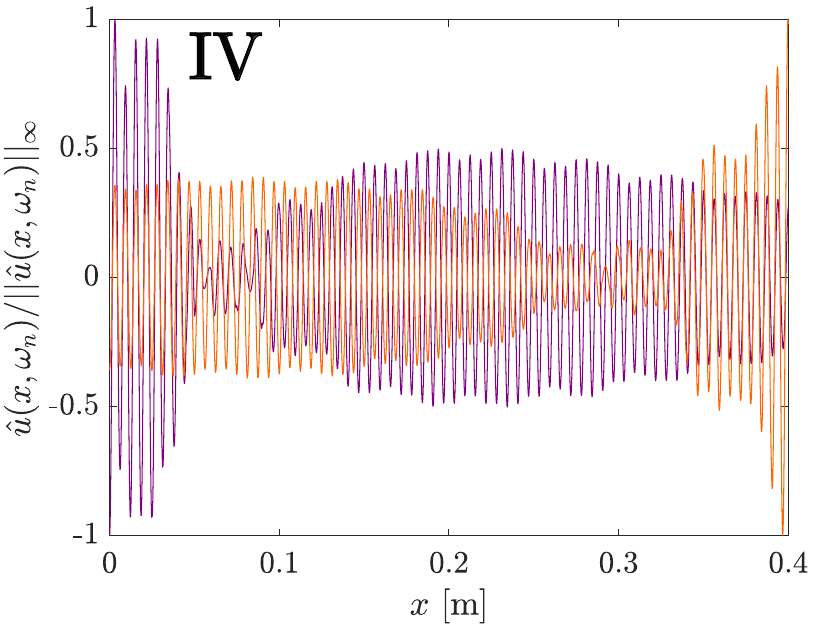}\label{6f}}
\caption{\textbf{Dispersion topology and NH skin effect with local feedback interactions ($a = 0$)}. (a) The complex frequency plane obtained as a projection of the dispersion relation from the first four PBs with $\kappa_g = 1.5$. Shaded blue and red areas represent regions with winding number $\nu = −1$ and $\nu = 1$, respectively.(b)  The same is done for negative feedback $\kappa_g = -2$. Eigenfrequencies of a finite structure are displayed as black dots in (a). Selected eigenmodes were shown for (c) 1 PB,(d) 2 PB,(e) 3 PB and (f) 4 PB. Solid purple lines are used for negative feedback, whereas orange ones are used for positive feedback. The same colors mark the selected eigenfrequencies on (a) and (b).}
\label{fig:6}
\end{figure}

\raggedbottom
Next, we illustrate the NH skin effect that occurs for finite structures, manifesting itself as a localization of most of the bulk modes at one of the  boundaries~\cite{kunst2018biorthogonal,rosa2020dynamics,lee2019anatomy,gong2018topological}. The presence of skin modes can be predicted based on the topology of the complex dispersion bands. As the wavenumber evolves within a BZ, the corresponding dispersion band makes a path, which can be projected on the complex plane. Thus, the winding numbers are interpreted geometrically by simply counting the number of times the loop of each PB encircles while evolving along the first BZ. For non-trivial bands, the projection results in closed paths, whose topology is associated with non-zero winding number $\nu$ inside the PB. This is illustrated for both positive and negative feedback in Fig.~\ref{fig:6}, which displays the dispersion loops whose directions are indicated by the arrows. Shaded blue and red areas represent regions with winding number $\nu = - 1$ and $\nu = 1$, respectively. Similar to the behavior regarding amplification and attenuation, we note that switching the sign of the feedback gain produces a topological change in each band evidenced by the switching of their invariant $\nu$. 

Free vibration modes of the finite structure are computed via FEM by solving the corresponding eigenvalue problem. Some of the eigenfrequencies are displayed as black dots, and we have verified that all of them fall inside one of the dispersion loops. Selected eigenfrequencies on the four first PB are marked with orange (positive feedback) and purple (negative feedback) crosses. The corresponding real part of the eigenmodes are also displayed, normalized by their maximum absolute value.

The results are displayed on Fig.~\ref{6c}, Fig.~\ref{6d}, Fig.~\ref{6e}, and Fig.~\ref{6f} for the first, second, third, and fourth PB, respectively, with orange solid
lines for positive feedback and purple solid lines for negative feedback. The plots show the normalized real part of the eigenmodes, where $\omega_n$ is the real part of the corresponding eigenfrequency. Aligned with similar observations made with discrete lattice systems \cite{rosa2020dynamics}, the results confirm that modes belonging to regions with $\nu=1$ are localized at the right boundary of the structure, and with $\nu=-1$ at the left. 

We note that, while the dispersion bands are characterized by attenuation and amplification, all the eigenfrequencies of the considered finite structures have a small positive imaginary frequency component due to damping. This illustrates that finite domains of the feedback actuated waveguides may be stable even though infinite domains would not be so, since the waves would amplify indefinitely. 

\begin{figure}[H]
    \centering
    \includegraphics[width=\textwidth]{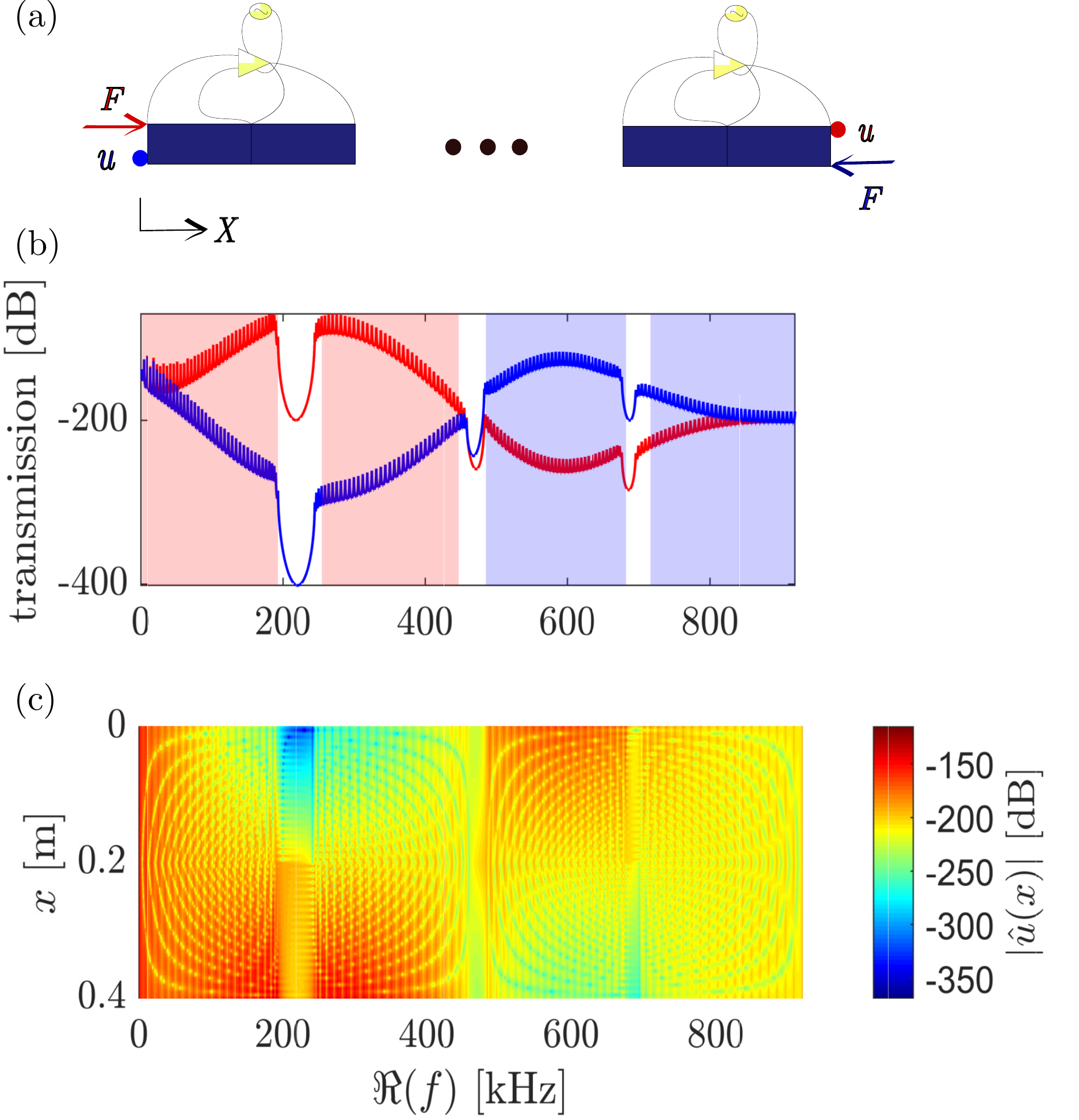}
    \caption{\textbf{Frequency responses of the structure in terms of displacement exhibiting energy concentration and non-reciprocal behavior - local feedback interactions.}(a) Scheme of the transmission simulation depicted on (b). (b) Results of displacement spectrum estimated at the right extremity $|\hat{u}(L_c)|$ with left extremity excitation (red lines) and at the left extremity $|\hat{u}(0)|$ with right extremity excitation (blue lines). (c) Two-dimensional plot of the harmonic response $|\hat{u}(x)|$ - in dB - with excitation at the middle of the structure ($x = 0.2$m).}
    \label{fig:10}
\end{figure}

Non-reciprocity and the localized modes arising from the NH skin effect are further investigated in terms of harmonic excitation. The frequency response functions (FRFs) of a finite structure are computed through the SEM approach. Fig.~\ref{fig:10}b displays the results for the negative feedback case ($\kappa_g = -2$ and $a = 0$), comparing the transmission from left-to-right (blue) to the transmission from right-to-left (red), as illustrated in the schematic of Fig.~\ref{fig:10}a. In the figure, shaded red and blue areas correspond to the bands previously identified to have $\nu=1$ and $\nu=-1$, respectively. This analysis highlights the strong non-reciprocal behavior of the waveguide for all the Pass bands: the bands with $\nu=1$ (red) tend to localize energy at the right boundary, while bands with $\nu=-1$ (blue) tend to localize at the left boundary. The behavior is further illustrated in Fig.~\ref{fig:10}c, which shows the response of the waveguide as a function of both frequency and position, when excited at its center. The figure clearly shows the tendency of localization at the right boundary for frequencies belonging to the shaded red bands ($\nu=1$), and to the left boundary for the shaded blue bands ($\nu=-1$)

\section{Non-local Feedback}

\begin{figure}[H]
\centering
\subfigure[]{\includegraphics[width=0.495\textwidth]{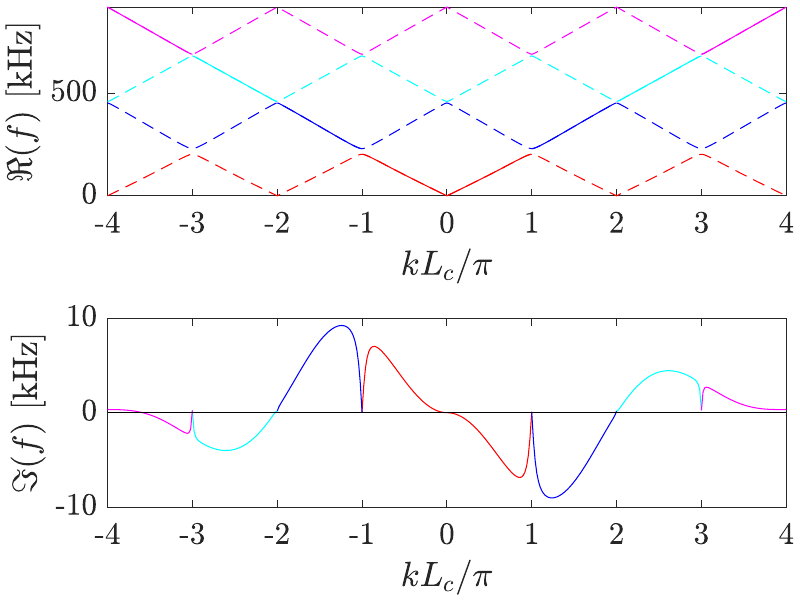}\label{7a}}
\subfigure[]{\includegraphics[width =0.495\textwidth]{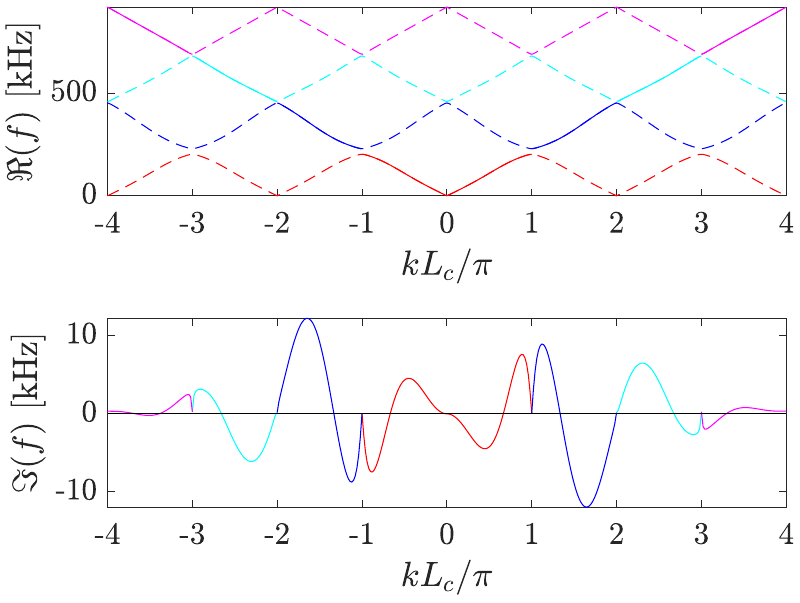}\label{7b}}
\caption{\textbf{Propagation modes of the dispersion diagrams of the piezoelectric material of Fig.~\ref{modelcont} compared for fixed $\kappa_g = -0.7$ while varying the locality parameter $a$.} (a) $a = 0$, and (b) $a = 1$. Each bulk band has its topology depicted with a different color: 1 PB is in solid red lines, 2 PB is in solid blue lines, 3 PB is in solid cyan lines and, finally, 4 PB is in solid magenta lines. On the real part, periodization of each band with respect to the wavenumber is represented by the dashed lines.}
\label{fig:7}
\end{figure}

We now discuss the wave propagation behavior using non-local feedback interactions ($a > 0$). Figure (\ref{fig:7}) compares the unfolded dispersion diagrams for the two different values of the locality parameter $a$ used in the main results of this work ($a=0$ and $a=1$). We note that each Pass bands of the local case ($a=0$) is split into two bands with opposite non-reciprocal behavior for the non-local case ($a=1$). 
This behavior was illustrated for discrete lattices in \cite{rosa2020dynamics} and is here confirmed within the framework of continuous waveguides with feedback interactions. In general, we observe that increasing the degree of non-locality $a$ results in the splitting of each original band (for $a=0$) into $a+1$ sub-bands with interchanging non-reciprocal behavior. For higher values of $a$, we refer the reader to  appendix. This feature seems to be interesting for the tuning of frequencies that are intended to propagate or not, with potential applications to wave filtering and multiplexing. 

\subsection{Non-reciprocal wave propagation}

\begin{figure}[H]
\centering
\subfigure[]{\includegraphics[width=0.495\textwidth]{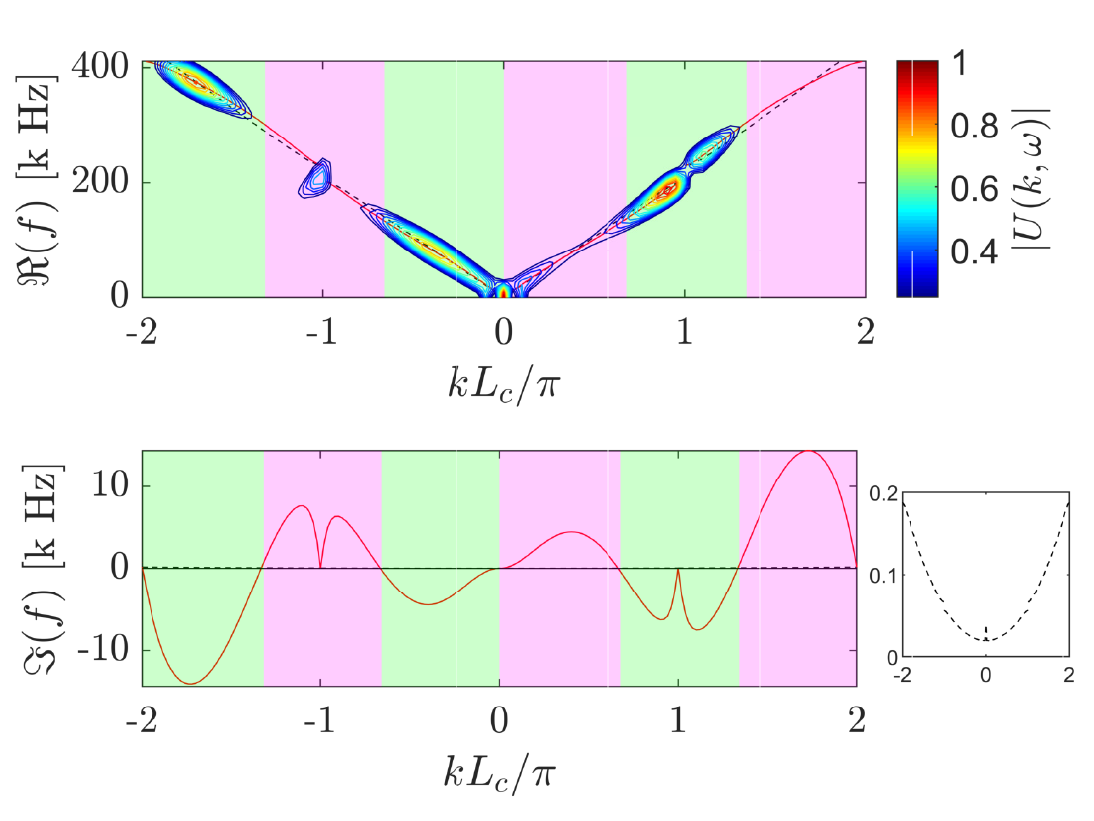}\label{8a}}
\subfigure[]{\includegraphics[width =0.495\textwidth]{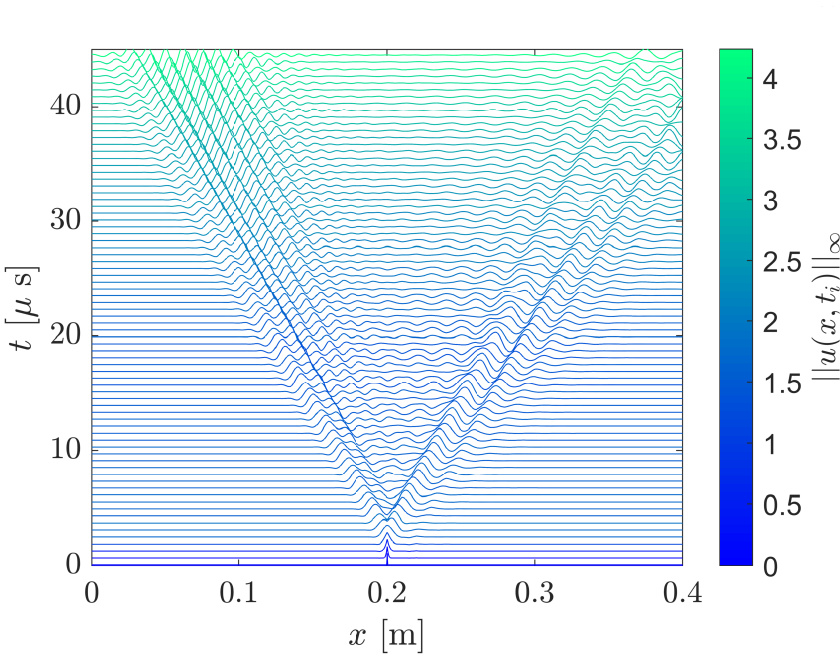}\label{8b}}
\subfigure[]{\includegraphics[width =0.495\textwidth]{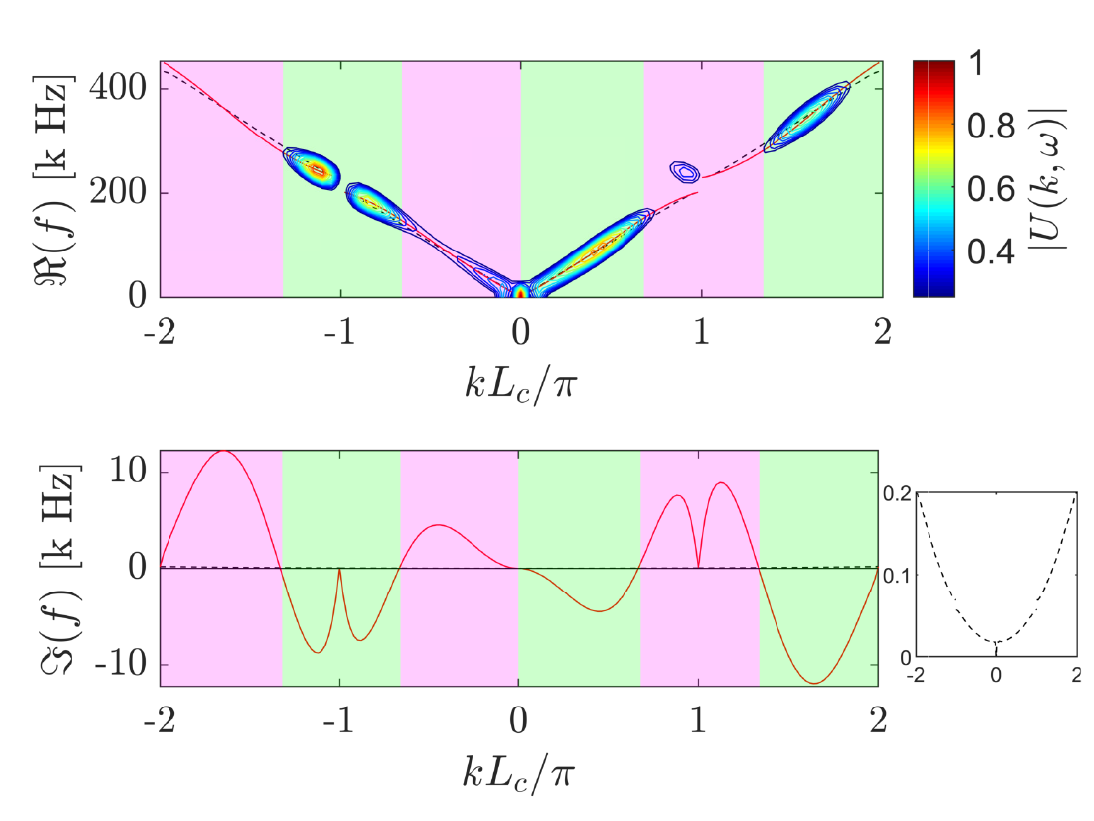}\label{8c}}
\subfigure[]{\includegraphics[width =0.495\textwidth]{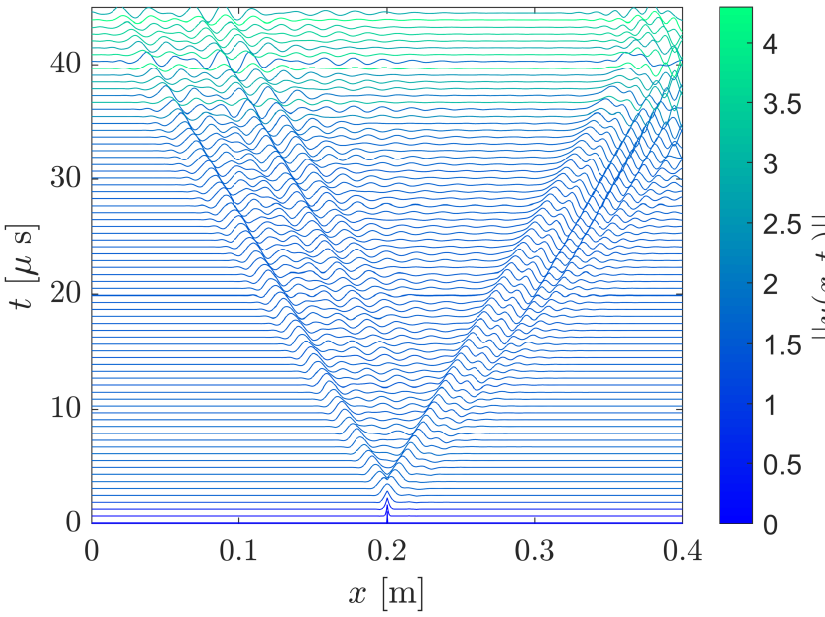}\label{8d}}
\caption{\textbf{Non-reciprocal amplification and attenuation of waves in structures with non-local feedback interactions ($a = 1$).} (a) and (c) Propagation modes of the dispersion relation for $\kappa_g = 0.8$ and $\kappa_g = -0.7$, respectively. First and Second PB are shown in solid red lines, superimposed to results for the equivalent passive lattice (making $\kappa_g = 0$) displayed in dashed black lines. The inlets show zoomed-in plots of the imaginary dispersion diagrams to allow visualizing the small amplitude reciprocal results for damped passive lattices. Attenuation and amplification zones are identified, respectively, by shaded
magenta and green areas. (b) and (d) Transient responses illustrating the non-reciprocal wave propagation that results from this odd property for $\kappa_g = 0.8$ and $\kappa_g = -0.7$, respectively. Non-reciprocal wave propagation is further confirmed by their dispersion estimation through 2DFFT displayed as contours in (a) and (c) which are normalized by their maximum value.% (e) and (f) Zoom on the imaginary dispersion diagrams to show the small and reciprocal results for damped passive lattices.
}
\label{fig:8}
\end{figure}

The same procedure described before was conducted for the non-local waveguides with $a=1$, but now employing a broadband spectrum excitation consisting on a tone-burst type signal with $1$ cycle and central frequency of $200$ [kHz], aiming at exciting several Pass bands at once. The results of the transient simulations are summarized in Fig.~\ref{fig:8} for both negative and positive feedback gains. The results illustrate once again the amplification along opposite directions for each Pass bands, their overall inversion when the sign of the feedback is switched, and the agreement with numerical dispersion relations confirmed by the contours of the 2D FFTs. In particular, we observe how indeed each Pass bands defined previously for $a=0$ now defines two bands with opposite directions for amplification. This general feature is further illustrated in the appendix for $a=2$ and $a=5$, where the splitting of the spectrum into multiple non-reciprocal bands is observed. 

\subsection{Skin modes and localized vibration}

\begin{figure}[H]
\centering
\subfigure[]{\includegraphics[width=0.495\textwidth,height=0.25\textheight]{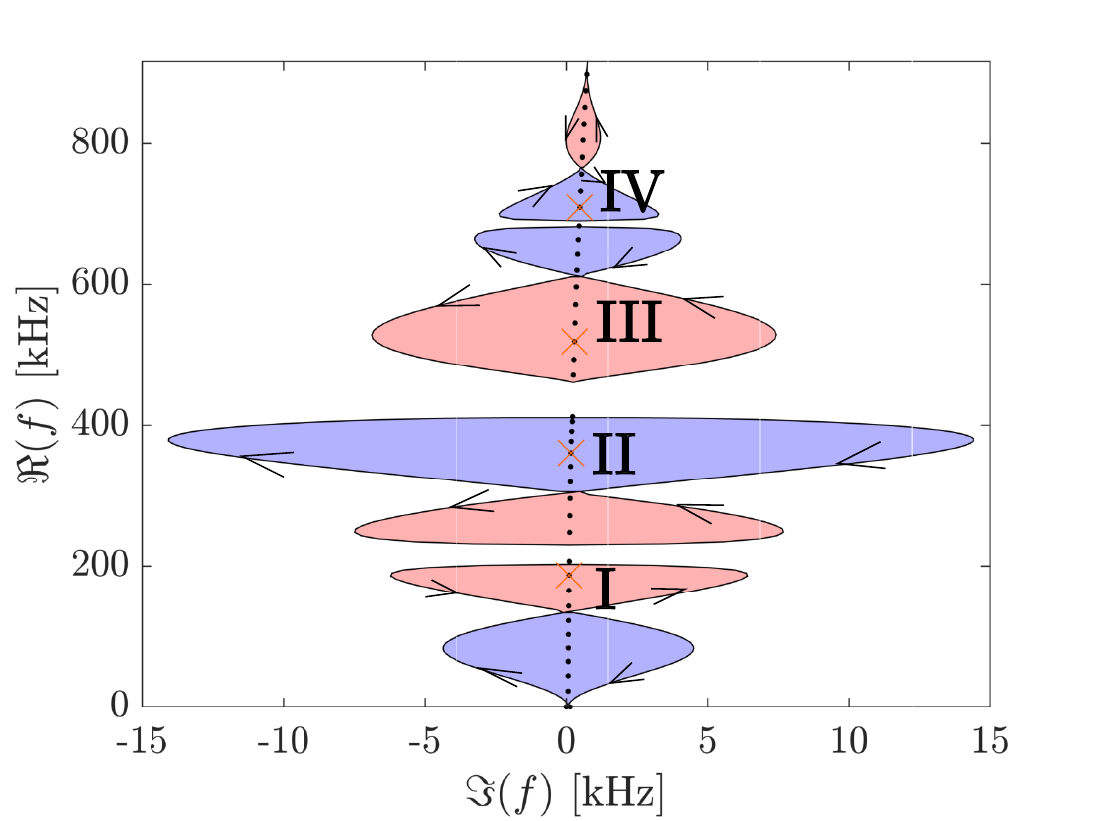}\label{9a}}
\subfigure[]{\includegraphics[width=0.495\textwidth,height=0.25\textheight]{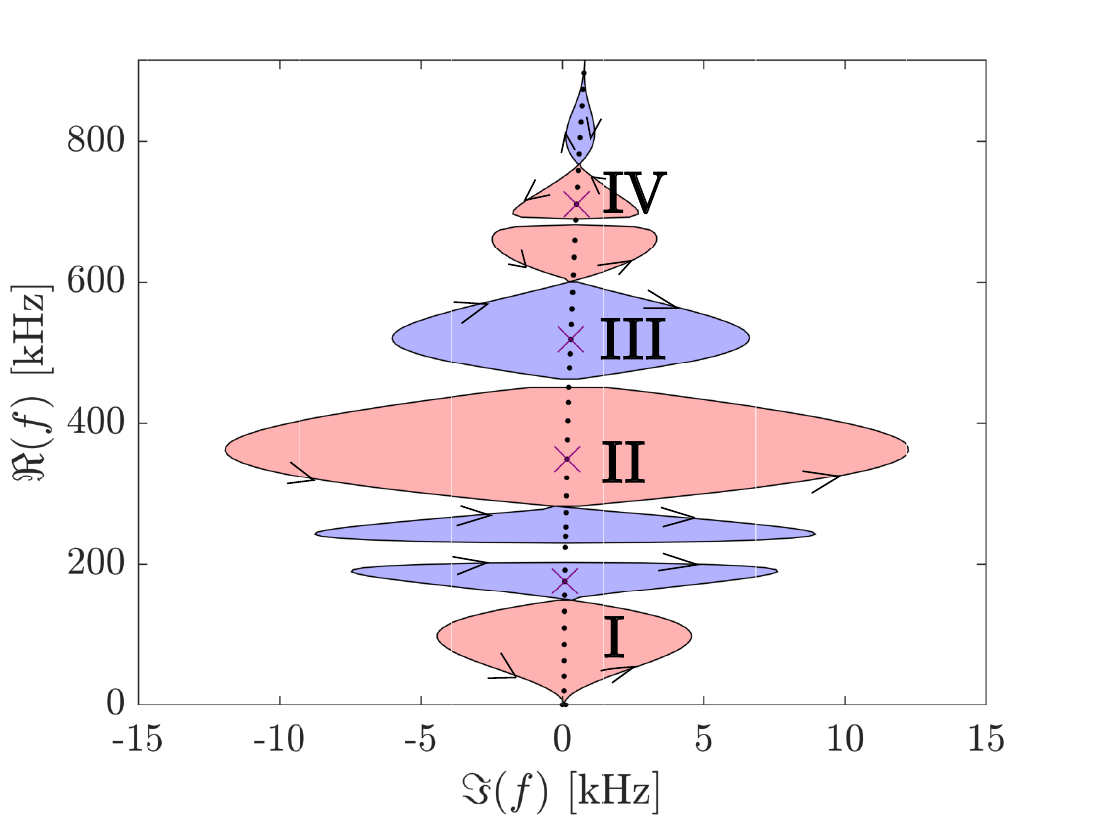}\label{9b}}
\subfigure[]{\includegraphics[width=0.495\textwidth,height=0.25\textheight]{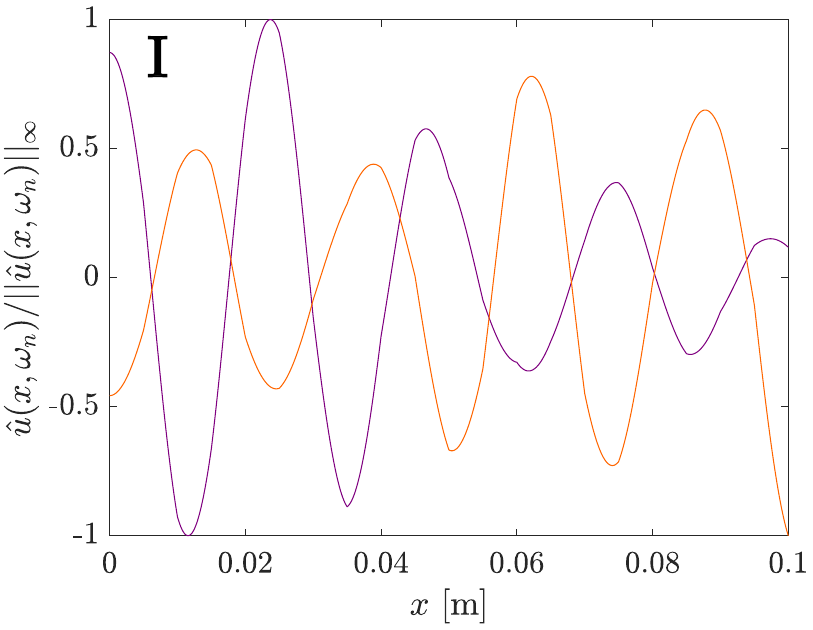}\label{9c}}
\subfigure[]{\includegraphics[width=0.495\textwidth,height=0.25\textheight]{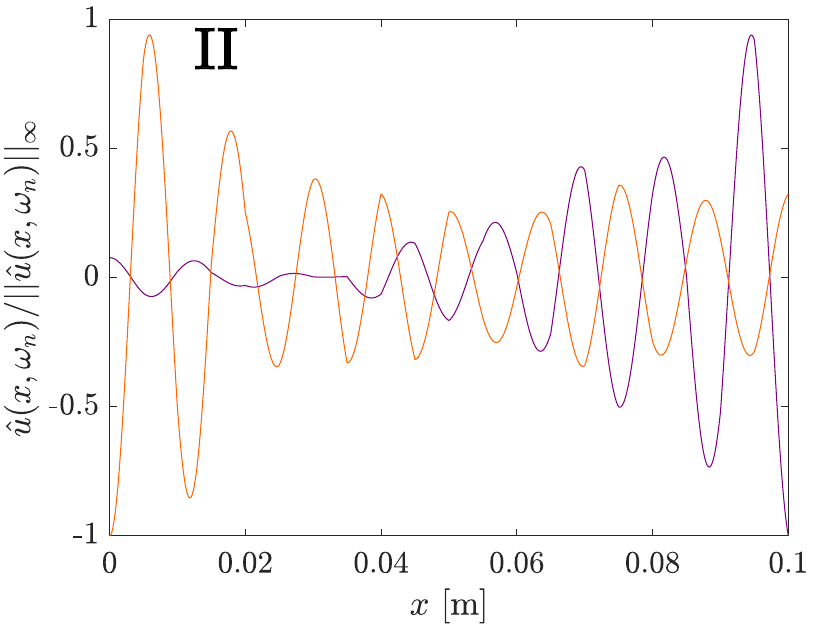}\label{9d}}
\subfigure[]{\includegraphics[width=0.495\textwidth,height=0.25\textheight]{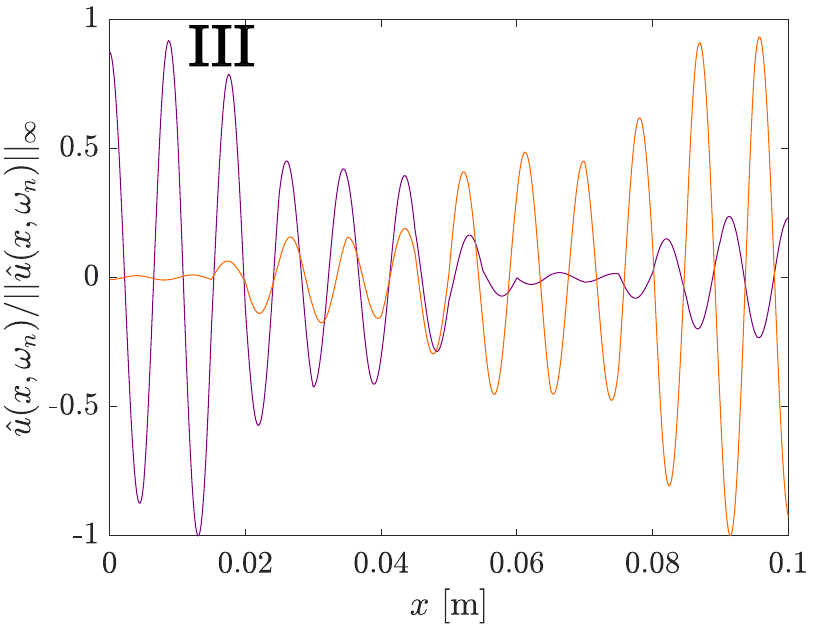}\label{9e}}
\subfigure[]{\includegraphics[width=0.495\textwidth,height=0.25\textheight]{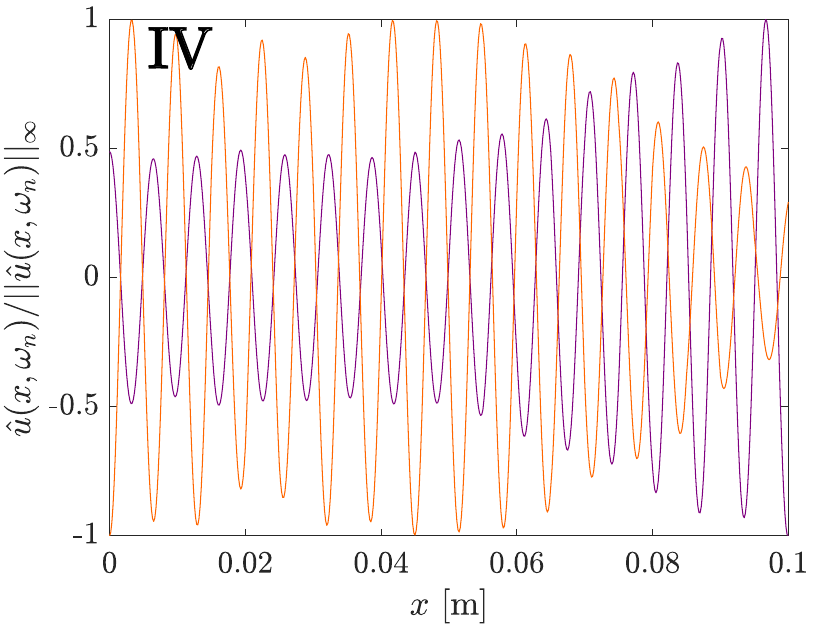}\label{9f}}
\caption{\textbf{Dispersion topology and NH skin effect with non-local feedback interactions ($a = 1$)}. (a) The complex frequency plane of the dispersion relation for the first four PBs with $\kappa_g = 0.8$. Shaded blue and red areas represent regions with winding number $\nu = −1$ and $\nu = 1$, respectively.(b) The same is done for negative feedback $\kappa_g = -0.7$. Eigenfrequencies of a finite structure are displayed as black dots in (a). Selected eigenmodes were taken for (c) 1 PB,(d) 2 PB,(e) 3 PB and (f) 4 PB. Solid purple lines were used for negative feedback whereas the orange ones were used for positive feedback. The same colors mark the selected eigenfrequencies on (a) and (b).}
\label{fig:9}
\end{figure}

As previously mentioned, the topology of the bands get altered by increasing the degree of non-locality through the parameter $a$. This is illustrated on Fig.~\ref{fig:9}, which displays the complex dispersion loops and eigenfrequencies of the finite domain for $a=1$ and both positive and negative feedback gains. We note that each dispersion loop previously observed for $a=0$ gets split into two dispersion loops with different winding numbers $\nu$. Also, again the topology of the bands gets inverted when switching the sign of the feedback gain $k_g$. The pictures also show the normalized real part of the eigenmodes, where we again confirm that eigenmodes of the finite domain are localized at the right boundary when their eigenfrequencies lie in regions of the complex plane with $\nu=1$, and at the left boundary for $\nu=-1$. 

\begin{figure}[H]
    \centering
    \includegraphics[width=\textwidth]{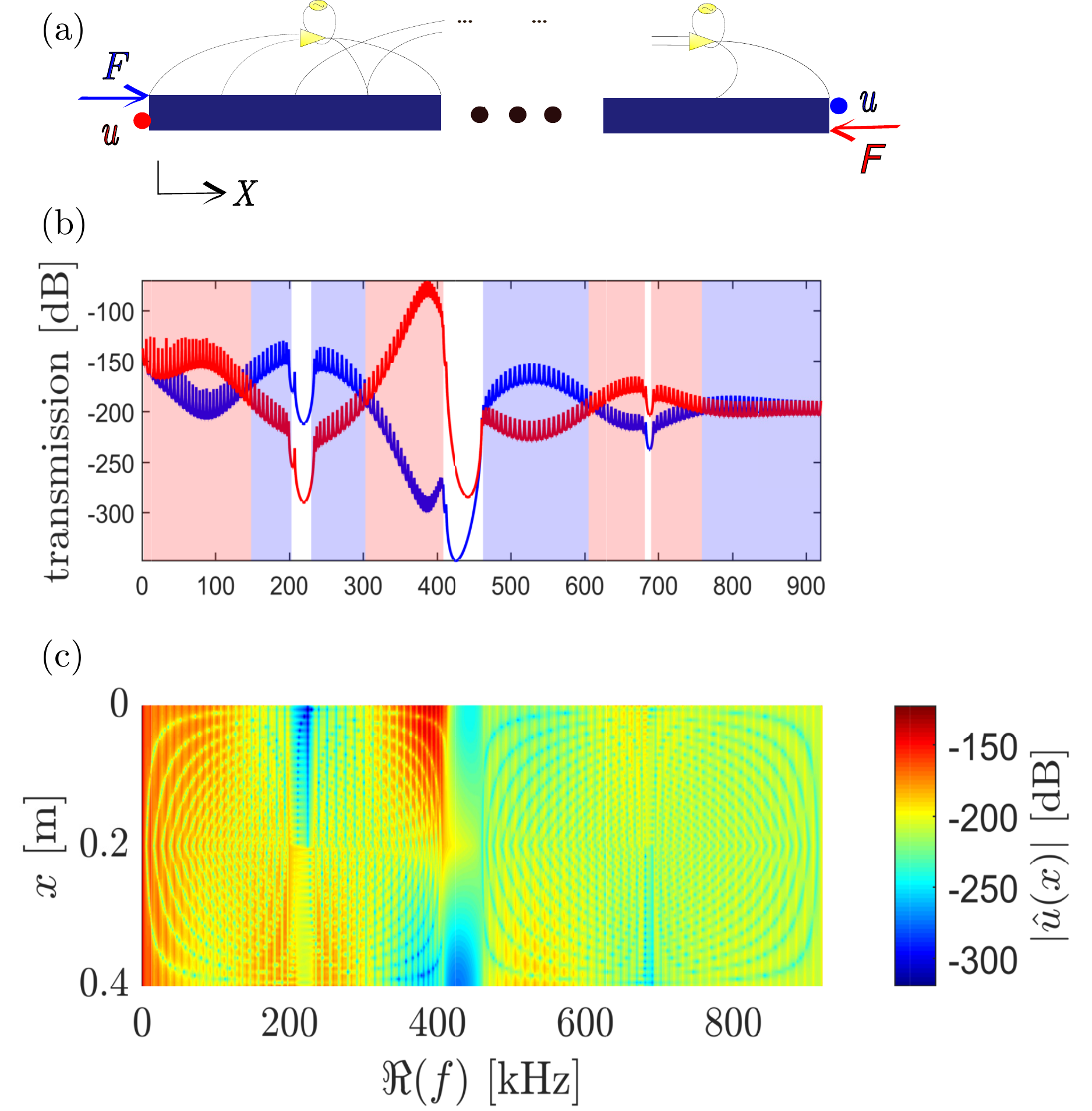}
    \caption{\textbf{Frequency responses of the structure exhibiting energy concentration and non-reciprocal behavior - non-local feedback interactions.}(a) Scheme of the transmission simulation depicted on (b). (b) Results of displacement spectrum estimated on the right extremity $|\hat{u}(L_c)|$ with left extremity excitation (red lines) and on the left extremity $|\hat{u}(0)|$ with right extremity excitation (blue lines). (c) Two-dimensional plot of $|\hat{u}(x)|$ - in dB - with excitation at the middle of the structure ($x = 0.2$m).}
    \label{fig:11}
\end{figure}

Figure~\ref{fig:11} illustrates the harmonic response of the structure using the same procedures as in Fig.~\ref{fig:10}. The numerical results were simulated for positive feedback of a non-local structure ($\kappa_g = 0.8$ and $a = 1$). Again we notice the strong non-reciprocity in terms of left-to-right and right-to-left transition, which is in agreement with the highlighted non-reciprocal bands. We also verify the localization at the boundaries according to the non-Hermitian skin effect when the structure is excited at its middle. Naturally, the non-reciprocity and localization are stronger on the first few PBs since their dispersion is characterized by larger imaginary frequency components. Although some bands have lower localization and non-reciprocal behavior, these results confirm the splitting of the bands into multiple sub-bands with opposite behavior due to the non-locality parameter $a=1$. 

%\textcolor{red}{this figure was wrong. We wrongly put Fig. 4 here on the sent manuscript. Now the Figure is the correct correspondence to non-local feedback with $a=1$.}
% esse tipo de comentário deve estar só na carta de resposta

\section{Conclusions}

In this work we investigated non-Hermitian elastic waveguides with feedback interactions realized through piezoelectric materials. Numerical results illustrate the non-reciprocity emanating from local and non-local feedback interactions, both in terms of wave propagation and localized modes of finite structures. These results expand previous observations on discrete lattice systems \cite{rosa2020dynamics} to distributed parameter structural models, in particular within a platform that can be built and tested in future experimental studies.

In addition to experimental investigations, future work may explore two-dimensional systems, other types of feedback interactions and the effects of non-linear components. The stability of this type of systems is also something to be explored in more detail, as it may impose limitations when conducting experimental investigations.

\section*{Acknowledgments}

The authors thank the financial support provided by the S\~ao Paulo Research Foundation (FAPESP) grant numbers \#2018/15894-0, \#2019/20235-9, and \#2020/07703-0. M. I. N. Rosa would like to thank the support from the National Science Foundation (NSF) through the EFRI 1741685 grant and from the Army Research Office (ARO) through grant W911NF-18-1-0036. The authors also thank Professor Massimo Ruzzene for the fruitful and insightful technical discussions.

\appendix
\section{Appendix}
\subsection{Numerical models for one-dimensional piezoelectric material under longitudinal vibration}
\label{A1: FEM_n_SEM}
\subsubsection{Finite Element Method (FEM)}

Consider the one-dimensional phononic crystal based on the unit cell displayed on Fig.~\ref{modelcont}. Let the $x$-axis be the longitudinal direction of the cell and, in addition, let the piezoelectric material (ceramics) be plated with electrodes at the end cross-sections. In this regard, $\sigma$ and $\varepsilon$ denote, respectively, the normal stress and normal strain, while $E$ and $D$ denote, respectively, the scalar electric field and the electric displacement in the same direction. 
% faz sentido dizer "normal direction" para campo elétrico? normal a que? 
Also, $Y$ is used herein referring to the corresponding Young's modulus, $\alpha$ denoting the relative permittivity of the medium %(sometimes called dielectric constant) 
and $e$ the piezoelectric constant. Therefore, the constitutive relation becomes \cite{won1992application}:

\begin{equation}
    \begin{cases}
    \sigma = Y \varepsilon - e E \\
    D = e \varepsilon + \alpha E,
    \end{cases}
    \label{constitutive_1}
\end{equation}

which can be integrated over the cross-section area by applying $\varoiint_A \,dA$ to both equations and recalling the definitions $N(x,t) = \varoiint_A \sigma(x,t) \, dA$ as the axial internal effort distribution along $x$ %the spatial domain 
and varying with time, $\varepsilon (t) = \frac{\partial u (x,t)}{\partial x}$, and $Q (x,t)= \varoiint_A D (x,t) \, dA$ the electric charge distribution.

It is assumed %the reasonable hypothesis 
that the electric field and electric displacement are uniform inside the piezoelectric material \cite{won1992application} so that $E(x,t) =E(t)$ and $D(x,t) = D(t).$ %We consider 
It is also assumed that $\sigma(x,t)$, $E(t)$ and $D(t)$ do not vary within the cross-section and that %. At last, 
the material properties are constant. With these assumptions, %is in mind, 
one can derive the constitutive equations for the internal force and electric charge \cite{zhao2009control}:

\begin{equation}
    \begin{cases}
    N = Y \varepsilon A - e A E  \\
    Q = e \varepsilon A + \alpha A E.
    \end{cases}
    \label{constitutive_2}
\end{equation}

Insofar as it is desirable to model the structure under known applied forces and electric voltages, one shall write the equations with $\varepsilon$ and $D$ as the independent variables, which can be achieved by the following rearrangement of (\ref{constitutive_1}):

\begin{equation}
    \begin{cases}
    \sigma = \left (Y + \frac{e^2}{\alpha} \right )\varepsilon - \frac{e}{\alpha} D \\
    E = -\frac{e}{\alpha} \varepsilon + \frac{1}{\alpha} D.
    \end{cases}
    \label{cosntitutive_3}
\end{equation}
Then, by applying the integration over the cross-sectional area on the first equation of (\ref{cosntitutive_3}),

\begin{equation*}
    N = \left (Y + \frac{e^2}{\alpha} \right )\frac{\partial u}{\partial x} A- \frac{e}{\alpha} Q.
\end{equation*}
%Hereafter, the subscript $(.)_e$ refers to each finite element, the subscript $(.)_s$ refers to the piezoelectric sensor variable, while the subscript $(.)_a$ refers to the piezoelectric actuator variables.

Using the second equation of (\ref{cosntitutive_3}) and the definition of voltage difference $V$ between two boundaries
spaced by length ($L$): 

\begin{equation}
     V = - \int_{0}^{L} E dx = \frac{e}{\alpha} \left [u(L) -u(0) \right ]- \frac{L}{\alpha A} Q,
     \label{voltage}
\end{equation}

which can be rewritten to give the expression of $Q$ - the charge related to the voltage difference $V$- as follows:

\begin{equation}
      Q  = \frac{e A}{L} \begin{bmatrix} -1 & 1\end{bmatrix}
      \begin{bmatrix}
      u(0)\\
      u(L)
      \end{bmatrix}
      -
      \frac{\alpha A}{L} V.
      \label{Q}
\end{equation}
 Hereafter, the subscript $(\;)_e$ in scalars and vectors and the superscript $(\;)^e$ in matrices refers to each finite element; the subscript $(\;)_s$ refers to the piezoelectric sensor variable, while the subscript $(\;)_a$ refers to the piezoelectric actuator variables. For the FEM formulation, consider that the displacement can be approximated by a linear function:

\begin{equation*}
    u(x,t) = \begin{bmatrix} 1 - \frac{x}{L_e} & \frac{x}{L_e}\end{bmatrix}
     \begin{bmatrix} u(0,t) \\ u(L,t)\end{bmatrix}_e \, .
\end{equation*}
Then, the strain is defined as

\begin{equation*}
    \varepsilon(x,t) = \begin{bmatrix} - \frac{1}{L_e} & \frac{1}{L_e}\end{bmatrix}
     \begin{bmatrix} u(0,t) \\ u(L,t)\end{bmatrix}_e \, .
\end{equation*}

It can be shown \cite{won1992application} that by applying a variational principle, one can derive the following equation, with $m$ being the mass per unity length and omitting time dependency of the notation:

\begin{equation}
   \frac{m L_e}{6} 
    \begin{bmatrix}
    2 & 1\\1 &2
    \end{bmatrix}
    \begin{bmatrix}
    \ddot{u}(0)\\ \ddot{u}(L)
    \end{bmatrix}_e
    + \left ( Y + \frac{e^2}{\alpha} \right ) \frac{A}{L_e}
     \begin{bmatrix}
    1 & -1\\-1 & 1
    \end{bmatrix}
    \begin{bmatrix}
    u(0)\\u(L)
    \end{bmatrix}_e
    -
    \frac{e}{\alpha} 
    \begin{bmatrix}
    -1\\
    1
    \end{bmatrix}
    Q
    =
    \begin{bmatrix}
    f(0)\\f(L)
    \end{bmatrix}_e,
    \label{FEM_Q}
\end{equation}
and with substitution of (\ref{Q}),

\begin{multline}
    \frac{m L_e}{6}
    \begin{bmatrix}
    2 & 1\\1 &2
    \end{bmatrix}
    \begin{bmatrix}
    \ddot{u}(0)\\ \ddot{u}(L)
    \end{bmatrix}_e
    + \left ( Y + \frac{e^2}{\alpha} \right ) \frac{A}{L_e}
     \begin{bmatrix}
    1 & -1\\-1 & 1
    \end{bmatrix}
    \begin{bmatrix}
    u(0)\\u(L)
    \end{bmatrix}_e
    -
    \frac{e^2 A}{\alpha L_a} 
   \begin{bmatrix} 
   1 & -1\\
   -1 & 1
   \end{bmatrix}
    \begin{bmatrix}
    u(0)\\u(L)
    \end{bmatrix}_a
    =  \\
    \\
    =  
    \begin{bmatrix}
    f(0)\\f(L)
    \end{bmatrix}_e
    -\frac{A}{\alpha L_a} 
    \begin{bmatrix}
    -1 \\ 1
    \end{bmatrix} V.
    \label{FEM_V}
\end{multline}

We define the well-known FEM mass matrix $\mathbf{M^e}$, and the piezoelectric stiffness matrix $\mathbf{K_p^e}$ as

\begin{equation*}
    \mathbf{M^e} =\frac{m L_e}{6}
    \begin{bmatrix}
    2 & 1\\1 &2
    \end{bmatrix},\hspace{0.5 cm}
    \mathbf{K_p^e} = \left ( Y + \frac{e^2}{\alpha} \right ) \frac{A}{L_e}
    \begin{bmatrix}
    1 & -1\\-1 & 1
    \end{bmatrix}.
\end{equation*}

Now, consider an electric open boundary condition for the %shunted circuit, as it is for the 
sensor part of the piezoelectric material, where $Q$ is zero. Then, from (\ref{FEM_Q}), the equation becomes

\begin{equation}
    \mathbf{M^e}
    \begin{bmatrix}
    \ddot{u}(0)\\ \ddot{u}(L)
    \end{bmatrix}_e
    +\mathbf{K_p^e}
    \begin{bmatrix}
    u(0)\\u(L)
    \end{bmatrix}_e
    =
    \begin{bmatrix}
    f(0)\\f(L)
    \end{bmatrix}_e.
    \label{FEM_sensor}
\end{equation}

While, for the actuator segment of the piezoelectric rod, the applied voltage is defined as a function of the sensor displacement as in (\ref{Kg}).

\begin{equation}
    V = K_g \left[ u(L) - u(0) \right ]_s 
    =
    K_g 
    \begin{bmatrix}
    -1 &  1
    \end{bmatrix}
    \begin{bmatrix}
    u(0)\\u(L)
    \end{bmatrix}_s \,
    \label{Kg}
\end{equation}

Finally, substitution in (\ref{FEM_V}) leads to

\begin{multline*}
    \frac{m L_e}{6}
    \begin{bmatrix}
    2 & 1\\1 &2
    \end{bmatrix}
    \begin{bmatrix}
    \ddot{u(0)}\\ \ddot{u(L)}
    \end{bmatrix}_e
    + \left ( Y + \frac{e^2}{\alpha} \right ) \frac{A}{L_e}
     \begin{bmatrix}
    1 & -1\\-1 & 1
    \end{bmatrix}
    \begin{bmatrix}
    u(0)\\u(L)
    \end{bmatrix}_e
    -
    \frac{e^2 A}{\alpha L_a} 
   \begin{bmatrix} 
   1 & -1\\
   -1 & 1
   \end{bmatrix}
    \begin{bmatrix}
    u(0)\\u(L)
    \end{bmatrix}_a
    =  \\
    \\
    = 
    \begin{bmatrix}
    f(0)\\f(L)
    \end{bmatrix}_e
    -
    K_g
    \frac{A}{\alpha L_a} 
    \begin{bmatrix}
    1 & -1 \\
    -1 & 1
    \end{bmatrix}
    \begin{bmatrix}
    u(0)\\u(L)
    \end{bmatrix}_s.
\end{multline*}

 We also define the actuator stiffness matrix $\mathbf{K_a}$ and the feedback matrix $\mathbf{\Gamma_c}$. 

\begin{equation*}
     K_a =  -\frac{e^2 A}{\alpha L_a} 
    \begin{bmatrix}
    1 & -1\\
    -1 & 1\\
    \end{bmatrix},\hspace{0.5 cm}
    \Gamma_c= K_g
    \frac{A}{\alpha L_a} 
    \begin{bmatrix}
    1 & -1 \\
    -1 & 1
    \end{bmatrix},
\end{equation*}

so we can rewrite

\begin{equation}
    \mathbf{M^e}
    \begin{bmatrix}
    \ddot{u}(0)\\ \ddot{u}(L)
    \end{bmatrix}_e
    +\mathbf{K_p^e}
    \begin{bmatrix}
    u(0)\\u(L)
    \end{bmatrix}_e
    +\mathbf{K_a}
    \begin{bmatrix}
    u(0)\\u(L)
    \end{bmatrix}_a
    +\mathbf{\Gamma_c}
    \begin{bmatrix}
    u(0)\\u(L)
    \end{bmatrix}_s
    =
    \begin{bmatrix}
    f(0)\\f(L)
    \end{bmatrix}_e.
    \label{FEM_actuator}
\end{equation}

\subsubsection{Spectral Element Solution (SEM)}

The constitutive equations (\ref{constitutive_2}) of the piezoelectric material can be written, in the frequency domain for each frequency $\omega$, recalling that $\varepsilon(x,t) = \frac{\partial u (x,t)}{\partial x}$ and $E(x,t) = - \frac{\partial \varphi (x,t)}{\partial x}$, as follows:

\begin{equation}
    \hat{N} (x,\omega) = Y A\frac{d \hat{u} (x, \omega)}{dx} + e A \frac{d \hat{\varphi} (x, \omega)}{dx},
    \label{constitutiveN}
\end{equation}

\begin{equation}
    \hat{Q} (x, \omega) = e A\frac{d \hat{u} (x, \omega)}{dx} - \alpha A \frac{d \hat{\varphi} (x, \omega)}{dx}.
    \label{constitutiveQ}
\end{equation}

The additional mechanical variable is the material mass density, $\rho$, whereas the geometric variable $A$ is the cross-section area. The electric domain has the variables $\varphi(x,t)$ as the electric potential field distributed along the $x$-axis. %As one can see, 
The piezoelectric material is a transducer which couples an elastic domain with a dielectric domain due to the piezoelectric effect.

Given that the general form of the differential equations of motion for the piezoelectric material can be expressed - with the double dot denoting second derivative with respect to time - as \cite{qian2004dispersion}:

\begin{equation*}
    \begin{cases}
    \rho A \ddot{u} = A\frac{\partial \sigma (x,t)}{\partial x} \\
    \frac{\partial D (x,t)}{\partial x} = 0,
    \end{cases}
\end{equation*}

one can substitute (\ref{constitutive_1}) and the definitions of normal strain and electric field, to obtain (\ref{motion}). 

\begin{equation}
    \begin{cases}
    \rho A \ddot{u} =  Y A \pdv[2]{ u (x,t)}{x} +e A \pdv[2]{ \varphi (x,t) }{x}  \\
    0 = e \pdv[2]{ u (x,t)}{x} - \alpha \pdv[2]{ \varphi (x,t) }{x}
    \end{cases}
    \label{motion}
\end{equation}

The second of the above equation shows that the electric voltage %function 
can be eliminated from the first equation, giving rise to the wave equation, which describes the one-dimensional longitudinal vibration within a piezoelectric uniform rod \cite{li2016analysis}, disregarding viscosity, as follows:

\begin{equation}
  \frac{\partial}{\partial x}\left[\left( Y + \frac{e}{\alpha} \right ) A \frac{\partial u (x,t)}{\partial x} \right] = \rho A \ddot{u} (x,t).
\label{eq: waveeq}
\end{equation}

Recalling (\ref{voltage}) and transforming from time to frequency domain, one obtain (\ref{eq star}). 

\begin{equation}
    \hat{V}(\omega) = \frac{e}{\alpha} \left[ \hat{u}(L,\omega) - \hat{u}(0,\omega) \right ] -\frac{\hat{Q}(\omega) L}{\alpha A}
    \label{eq star}
\end{equation}

Also, substituting (\ref{constitutiveQ}) in (\ref{constitutiveN}) results in

\begin{equation}
    \hat{N} (x,\omega) = \left (Y + \frac{e^2}{\alpha} \right ) A \frac{d \hat{u} (x, \omega)}{dx} - \frac{e}{\alpha} \hat{Q} (\omega).
    \label{eq starstar}
\end{equation}

Now, consider an electric open boundary condition for the sensor segment of the cell considered herein, with the right and left end displacements denoted by $u_s(0,t)$ and $u_s(L,t)$. This assumption implies $\hat{Q}(\omega) = 0$ and, hence, from (\ref{eq star}),

\begin{equation}
    \hat{V}_s(\omega) = \frac{e}{\alpha} \left[\hat{u}_s(L,\omega) - \hat{u_s}(0,\omega) \right],
    \label{Vs}
\end{equation}

and from (\ref{constitutiveQ}),

\begin{equation*}
    \frac{d \hat{\varphi}_s (x,\omega)}{dx} = \frac{e}{\alpha} \frac{d \hat{u} (x,\omega)}{dx},
\end{equation*}

Moreover, from (\ref{eq starstar}),

\begin{equation}
    \hat{N}_s(x,\omega) = \left ( Y + \frac{e^2}{\alpha} \right ) A \frac{d \hat{u} (x,\omega)}{dx}.
    \label{Nx_sensor_}
\end{equation}

 Imposing a wave solution of the form
 \begin{equation}
     \hat{u}(x,\omega) = a_1 e^{i k(\omega) x} + d_1 e^{-i k(\omega) x},
     \label{wave_sol}
 \end{equation}

 one can derive
  
 \begin{equation*}
     \frac{d\hat{u}(x,\omega)}{dx} = i k(\omega) \left ( a_1 e^{i k(\omega) x} - d_1 e^{-i k(\omega) x} \right ).
 \end{equation*}

 Here, $i$ denotes the imaginary number and $k$ the wavenumber.
 
  Moreover, it can be shown from  (\ref{eq: waveeq}) that the dispersion relation for the uniform piezoelectric rod is
 
 \begin{equation*}
     k = \omega \sqrt{\frac{\rho}{Y + \frac{e^2}{\alpha}}}.
 \end{equation*}

 The oscillatory terms mean waves propagating to the right and to the left or, in other words, an incident and a reflected wave.
 
 Applying this solution to the sensor domain $x \in [0,L_s] $, (\ref{Nx_sensor_}) can be rewritten as 
  
\begin{equation}
    \hat{N}_s(x, \omega) = \xi (\omega) \left( a_1 e^{i k (\omega) x} - d_1 e^{-i k(\omega) x} \right ) \hspace{0.5 cm} : \hspace{0.5 cm} \xi (\omega) = \left ( Y + \frac{e^2}{\alpha} \right ) A i k (\omega).
    \label{Nx_sensor}
\end{equation}

 On the other hand, if the electric boundary condition is an applied feedback gain of the form:
 
 \begin{equation*}
     \hat{V}_a(\omega)  = K_g \left[\hat{u}_s(L,\omega) - \hat{u}_s, (0,\omega)\right]
 \end{equation*}
 
  as is the case for the piezoelectric actuator, which has the  right and left end displacements denoted as $u_a(0,t)$ and $u_a(L,t)$, one can verify from (\ref{eq star}) that
 
 \begin{equation*}
     \hat{Q_a} (\omega)= \frac{e A}{L_a} \left[\hat{u}_a(L,\omega) -\hat{u}_a(0,\omega) \right] - K_g \frac{\alpha A}{L_a}\left[\hat{u}_s(L,\omega) -\hat{u}_s(0,\omega) \right] \, ,
 \end{equation*}

and from (\ref{constitutiveQ}),

\begin{equation*}
    \frac{d \hat{\varphi}_a (x,\omega)}{dx} = \frac{e}{\alpha} \frac{d \hat{u} (x,\omega)}{dx} + \frac{1}{L_a} K_g \left[\hat{u}_s(L,\omega) -\hat{u}_s(0,\omega) \right]  - \frac{e}{\alpha L_a}\left[\hat{u}_a(L,\omega) -\hat{u}_a(0,\omega) \right] \, ,
\end{equation*}

Moreover, from (\ref{eq starstar}),

\begin{equation*}
    \hat{N}_a (x,\omega) = \left ( E + \frac{e^2}{\alpha} \right ) A \frac{d \hat{u} (x,\omega)}{dx} + \frac{e A}{L_a} K_g \left[\hat{u}_s(L,\omega) -\hat{u}_s(0,\omega) \right] - \frac{e^2 A}{\alpha L_a}\left[\hat{u}_a(L,\omega) -\hat{u}_a(0,\omega) \right]. 
\end{equation*}

Again, considering the wave solution, this equation can be rewritten as 

\begin{equation}
    \hat{N}_a(x,\omega) = \xi \left( a_1 e^{i \kappa(\omega) x} - d_1 e^{-i \kappa(\omega) x} \right ) + \frac{e A}{L_a} K_g \left[\hat{u}_s(L,\omega) -\hat{u}_s(0,\omega) \right] - \frac{e^2 A}{\alpha L_a}\left[\hat{u}_a(L,\omega) -\hat{u}_a(0,\omega) \right].
    \label{Nx_actuator}
\end{equation}

For the derivation of the SEM dynamic stiffness matrix, one must use (\ref{wave_sol}) to write the boundary degrees of freedom in matrix form. Hereafter, frequency dependency will be omitted for the sake of notation simplicity. Applying the wave solution $\hat{u}(x) $ to the sensor domain, $x \in [0,L_s] $

\begin{equation*}
    \begin{bmatrix}
    \hat{u}(0)\\
    \hat{u}(L)\\
    \end{bmatrix}_s
    = \begin{bmatrix}
    1 &1\\ 
    e^{i k L_s} &e^{- i k L_s}
    \end{bmatrix}
    \begin{bmatrix}
    a_1\\
    d_1\\
    \end{bmatrix}_s,
\end{equation*}

from which comes

\begin{equation}
    \begin{bmatrix}
    a_1\\
    d_1\\
    \end{bmatrix}_s
    = \frac{1}{e^{- i k L_s} - e^{ i k L_s}}\begin{bmatrix}
    e^{- i k L_s} & -1\\ -e^{i k L_s} & 1
    \end{bmatrix}
    \begin{bmatrix}
    \hat{u}(0)\\
    \hat{u}(L)\\
    \end{bmatrix}_s.
    \label{aux}
\end{equation}

Additionally, using (\ref{Nx_sensor}), one can derive

\begin{equation*}
    \begin{bmatrix}
    \hat{N}(0)\\
    \hat{N}(L)\\
    \end{bmatrix}_s
    = \xi \begin{bmatrix}
    1 & -1\\ e^{i k L_s} & e^{- i k L_s}
    \end{bmatrix}
    \begin{bmatrix}
    a_1\\
    d_1\\
    \end{bmatrix}_s,
\end{equation*}

and using (\ref{aux}) we get (\ref{D_sensor}).

\begin{equation}
    \begin{bmatrix}
    \hat{f}(0)\\
    \hat{f}(L)\\
    \end{bmatrix}_s
    = K_p^s
    \begin{bmatrix}
    \hat{u}(0)\\
    \hat{u}(L)\\
    \end{bmatrix}_s
    \label{D_sensor}
\end{equation}

where

\begin{equation}
    K_p = \frac{- i \xi}{sen(k L)} \begin{bmatrix}
    cos(k L) & -1\\ -1 &  cos(k L)
    \end{bmatrix},
\end{equation}

and the external loads $\hat{f}(0) = -\hat{N}(0)$ and $\hat{f}(L) = \hat{N}(L)$ have taken the place of the internal reactions.

The same procedure must be applied to the actuator. Applying the wave solution $\hat{u}(x) $ to the actuator domain, $x \in [L_s,L_c] $ we find that

\begin{equation*}
    \begin{bmatrix}
    \hat{u}(0)\\
    \hat{u}(L)\\
    \end{bmatrix}_a
    = \begin{bmatrix}
    1 &1\\ e^{i k L_a} &e^{- i k L_a}
    \end{bmatrix}
    \begin{bmatrix}
    a_1\\
    d_1\\
    \end{bmatrix}_a.
\end{equation*}

Equation (\ref{Nx_actuator}) yields

\begin{equation*}
    \begin{bmatrix}
    \hat{N}(0)\\
    \hat{N}(L)\\
    \end{bmatrix}_a
    = \xi \begin{bmatrix}
    1 & -1\\ 
    e^{i k L_a} & e^{- i k L_a}
    \end{bmatrix}
    \begin{bmatrix}
    a_1\\
    d_1\\ 
    \end{bmatrix}
    - \frac{e^2 A}{\alpha L_a}
    \begin{bmatrix}
    -1 & 1\\
    -1 & 1\\
    \end{bmatrix}
    \begin{bmatrix}
    \hat{u}(0)\\
    \hat{u}(L)\\
    \end{bmatrix}_a
    + \frac{e A K_g}{L_a} 
    \begin{bmatrix}
    -1 & 1\\
    -1 & 1\\
    \end{bmatrix}
    \begin{bmatrix}
    \hat{u}(0)\\
    \hat{u}(L)\\
    \end{bmatrix}_s,
\end{equation*}

which turns out to be

\begin{equation}
    \begin{bmatrix}
    \hat{f}(0)\\
    \hat{f}(L)\\
    \end{bmatrix}_a
    = (\mathbf{K_p^a} + \mathbf{K_a})
    \begin{bmatrix}
    \hat{u}(0)\\
    \hat{u}(L)\\
    \end{bmatrix}_a
    + \mathbf{\Gamma_c} 
     \begin{bmatrix}
    \hat{u}(0)\\
    \hat{u}(L)\\
    \end{bmatrix}_s \,  .
    \label{D_actuator}
\end{equation}

Note that the short-circuited piezoelectric material has the frequency dependent dynamic stiffness matrix $\mathbf{K_{sc}} (\omega) = \mathbf{K_p^a} (\omega) + \mathbf{K_a}$, which is the passive contribution to the dynamic stiffness relation of interest. Naturally, $\mathbf{\Gamma_c}$ is the feedback interaction contribution. If the feedback gain $K_g = 0$, $\mathbf{\Gamma_c}$ becomes the null matrix, and the $\hat{V} = 0$, which is equivalent to a short-circuit boundary condition.

Concatenating the dynamic stiffness matrix from the sensor -  equation (\ref{D_sensor})- and the actuator segments - equation ( \ref{D_actuator}) -, one can derive the dynamic stiffness matrix of the unit cell, $\mathbf{D}(\omega)$:

\begin{equation}
        \mathbf{\hat{f}}
    = \mathbf{D}(\omega)
        \mathbf{\hat{u}}
\end{equation}

where $\mathbf{\hat{u}}$ is the vector with the degrees of freedom of the unit cell, while  $\mathbf{\hat{f}}$ is the vector with the corresponding external forces.

Assembling the cells into a finite metastructure, the spectral global equation is finally obtained as

\begin{equation}
        \mathbf{\hat{f}_g}
    = \mathbf{D_g}(\omega)
        \mathbf{\hat{u}_g},
\end{equation}

concatenating degrees of freedom and forces of the whole structure on the global vectors $\mathbf{\hat{u}_g}$ and $\mathbf{\hat{f}_g}$.

\subsection{Comparison between methods}
\label{A2: Comparing_FEM_n_SEM}

Aiming at comparing the mathematical models developed in previous section, i.e. SEM semi-analytical formulation and FEM approximation of a finite structure, we simulate the dispersion relations using both methods.

\begin{figure}[H]
\centering
\subfigure[]{\includegraphics[width=0.495\textwidth]{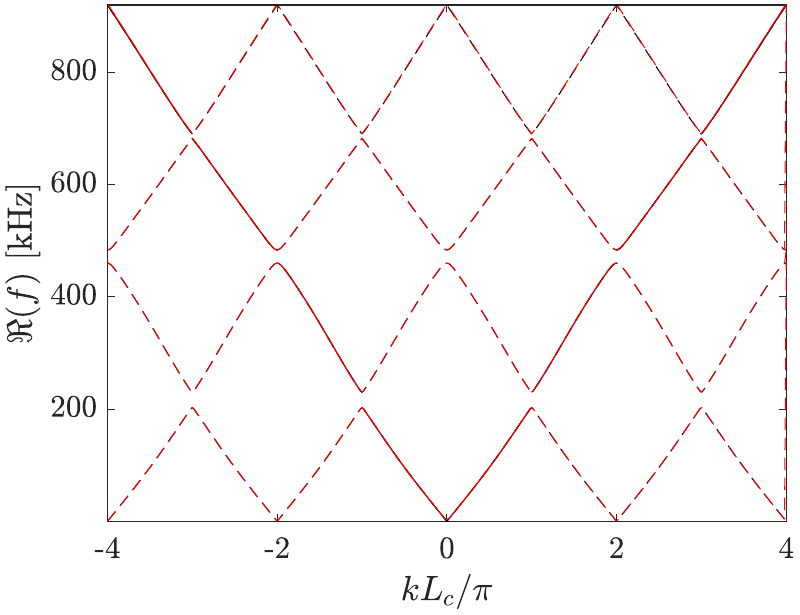}}
\subfigure[]{\includegraphics[width =0.495\textwidth]{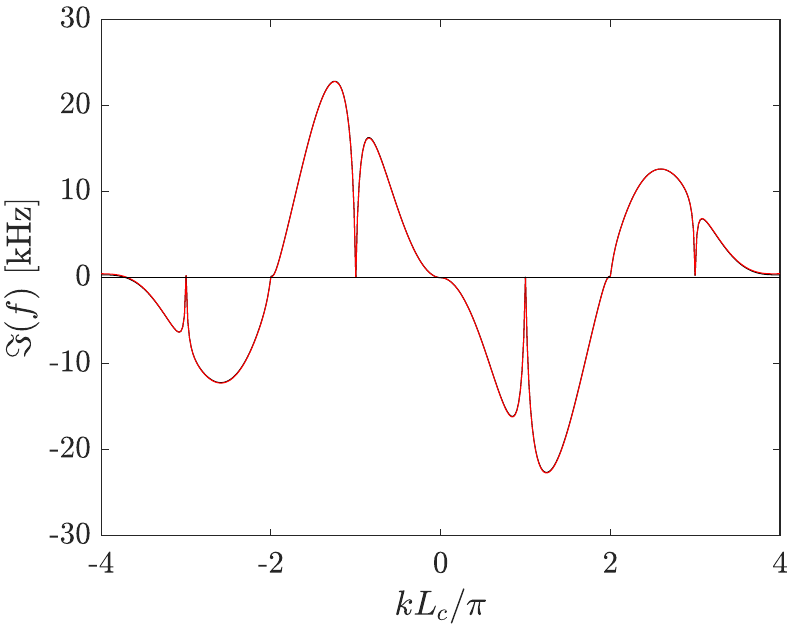}}
\caption{\textbf{Unfolded dispersion diagrams comparing SEM and FEM using $\kappa_g = -2 $ and $a = 0$.} Red lines represent the FEM approximation and black lines the SEM results.}
\label{SEM_FEM1}
\end{figure}

\begin{figure}[H]
\centering
\subfigure[]{\includegraphics[width=0.495\textwidth]{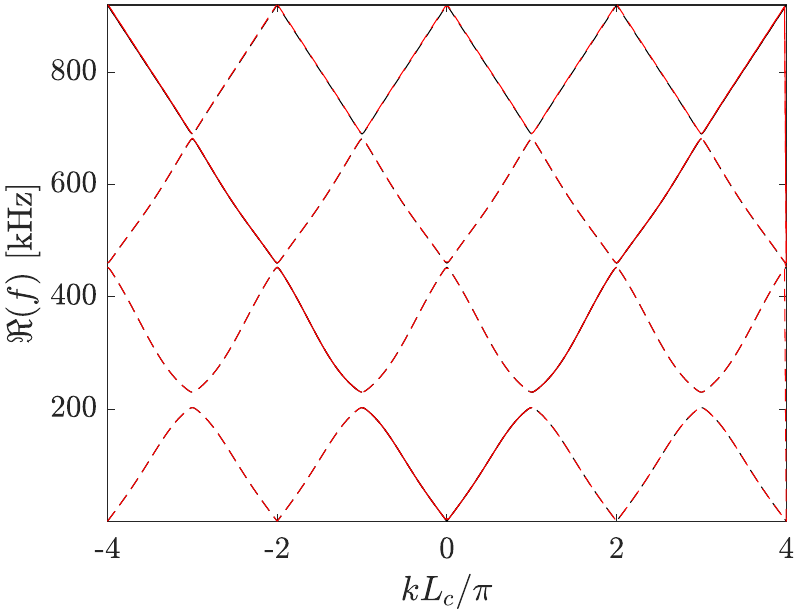}\label{2a_nonlocal}}
\subfigure[]{\includegraphics[width =0.495\textwidth]{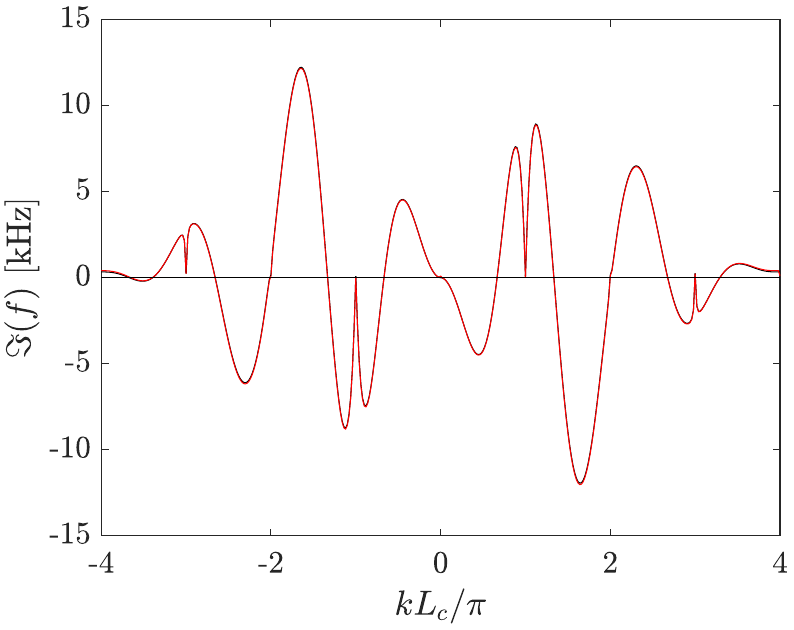}\label{2b_nonlocal}}
\caption{\textbf{Unfolded dispersion diagrams comparing SEM and FEM using $\kappa_g = -0.7 $ and $a = 1$.} Red lines represent the FEM approximation and black lines the SEM results.}
\label{SEM_FEM_2}
\end{figure}

The numerical results are displayed in Fig.~\ref{SEM_FEM1} for local and in Fig.~\ref{SEM_FEM_2} for non-local feedback interaction Solid black lines represent SEM solutions while solid red lines represent FEM approximate solutions. To realize asymptotically stable structural models, it is convenient % mandatory
to insert some source of damping. Rayleigh's viscous damping model was chosen for the FEM models, while in SEM models an imaginary part of the Young's modulus proportional to the loss factor $\eta$ was used as structural damping. Based on the work of \cite{hall2006problems}, a calibration function was used so that both models of damping give equivalent results.

\subsection{Folded and unfolded dispersion diagrams}
\label{A3: unfolded DD}

\begin{figure}[H]
\centering
\subfigure[]{\includegraphics[width=0.495\textwidth]{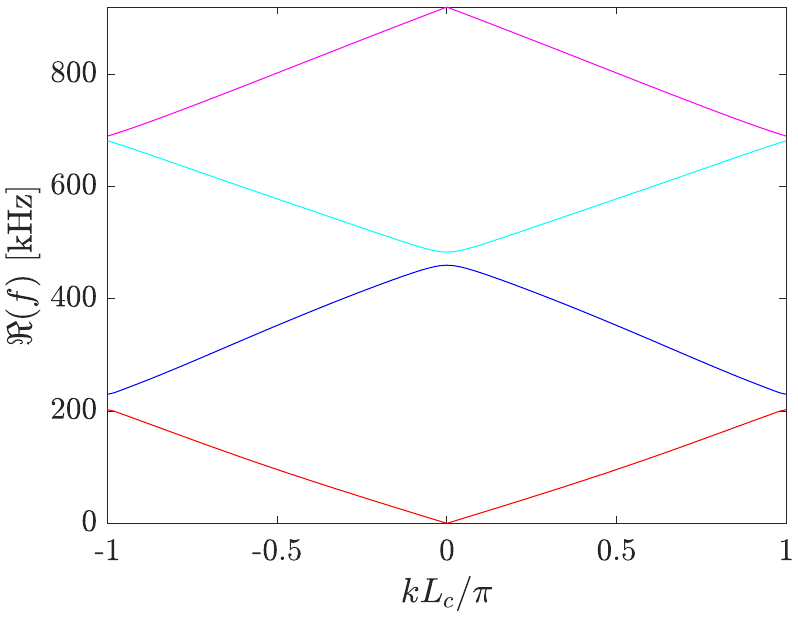}\label{2af}}
\subfigure[]{\includegraphics[width =0.495\textwidth]{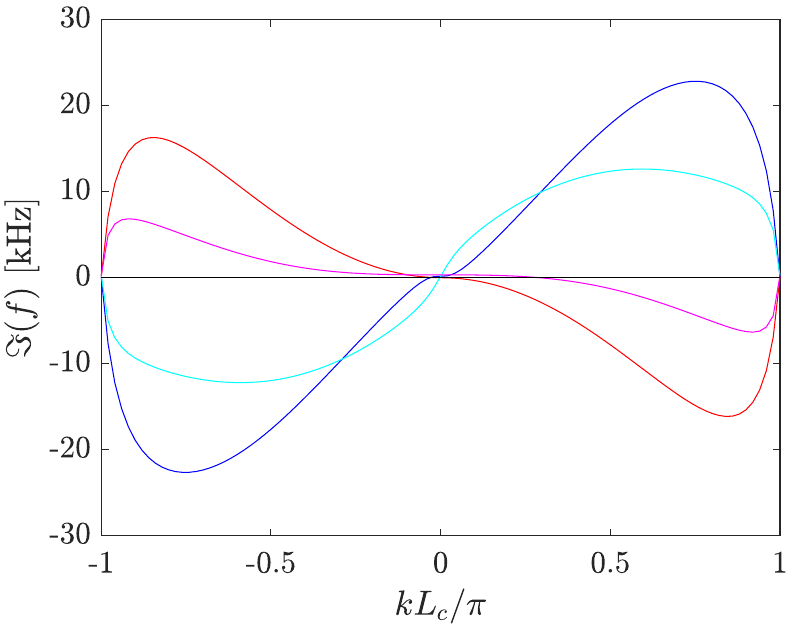}\label{2bf}}
\caption{\textbf{Folded dispersion diagrams from the cell of Fig.~\ref{modelcont}}. (a) Real part of the frequency versus the normalized wavenumber. (b) Imaginary part of the frequency versus the normalized wavenumber. Each color represents a different Pass bands.}
\label{foldedDD}
\end{figure}

In order to explain why the unfolded version of the dispersion diagram is necessary for the analysis of NH systems and avoiding falling into cumbersome explanations, we chose an illustrative argument. Consider the folded dispersion diagram depicted on Fig.~\ref{foldedDD} related to the same scheme simulated on Fig.~\ref{unfoldedDD}. Each color represents a Pass bands calculated via the inverse method~\cite{hussein2014dynamics}. The method relies on calculating $\omega (k)$ by imposing real values of the wavenumber, i.e., we are not interested in evanescent wave modes. These diagrams are said to be folded for the limitation they impose on $kL_c$ to assume values from $-\pi$ to $\pi$. Due to the periodicity of the relation, higher bulk bands are reflected into the first Brillouin zone (1BZ). If the intention is to analyze the topology of the diagrams - particularly what the imaginary frequency can tell us -, however, this procedure can lead to wrong conclusions. 
Consider the following remarks about the even Pass bandss (2 PB and 4 PB in the example): by Fig.~\ref{2bf}, the second Pass bands, represented by blue solid lines, apparently amplifies ($\operatorname{\mathfrak{I}}(f) < 0$) waves traveling from right to left ($ k < 0$), whereas the fourth Pass bands(magenta solid lines) seems to do the opposite, amplifying waves traveling from left to right ($k > 0$).

Nevertheless, the analysis for the unfolded dispersion diagrams, previously displayed in Fig.~\ref{unfoldedDD}, gave opposite conclusions, confirmed by the transient simulations presented in that same section and by the locality characterization, related to the skin modes, of each Pass bands. In fact, the second band amplifies waves traveling from left to right, whereas the fourth band amplifies waves traveling from right to left. 
As previously noticed, both 2 PB and 4 PB suffer inversion of their topology when reflected to the 1BZ. So, it is clear that conclusions about the wave propagation behavior based upon inverted bands, due to reflection on the 1BZ, are inconsistent.
 
\subsection{Non-locality effect}
\label{A4: nlocal}

\begin{figure}[H]
\centering
\subfigure[]{\includegraphics[width=0.495\textwidth]{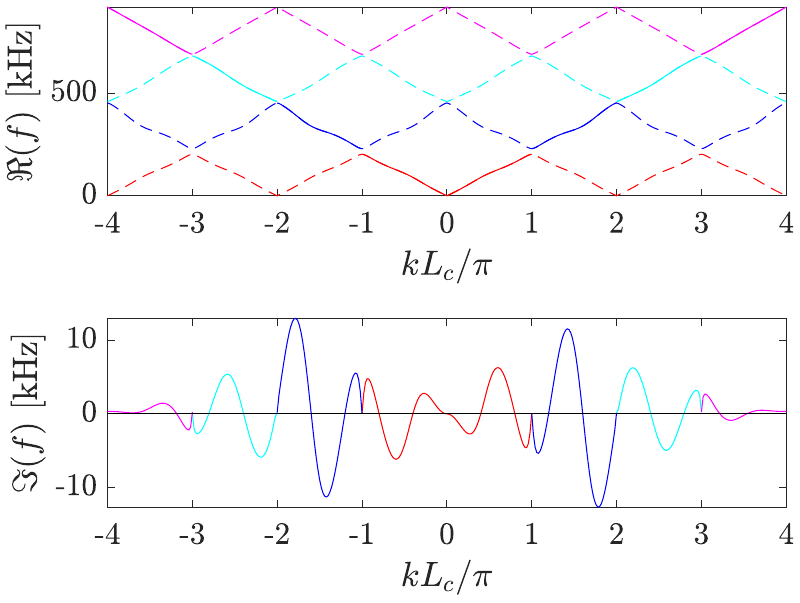}\label{14a}}
\subfigure[]{\includegraphics[width =0.495\textwidth]{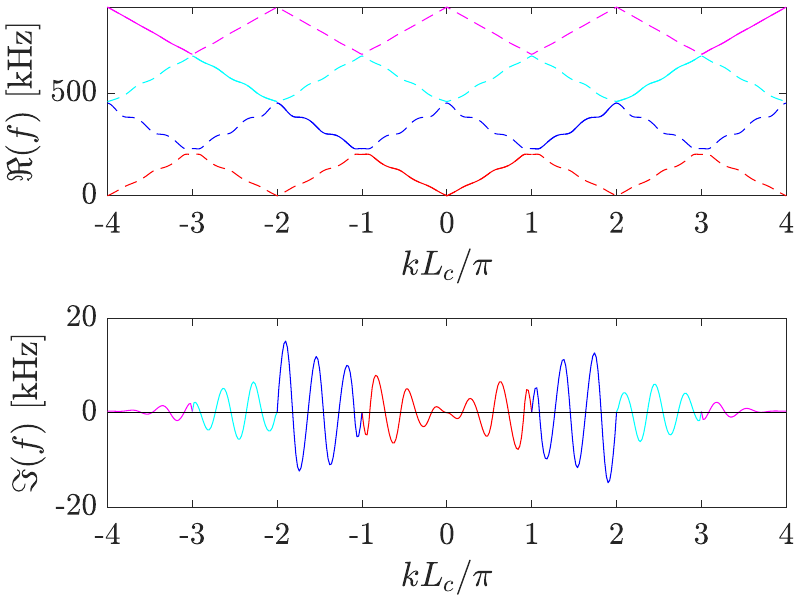}\label{14b}}
\caption{\textbf{Propagation modes of the dispersion diagrams of the piezoelectric material of Fig.~\ref{modelcont} compared for fixed feedback gain of $\kappa_g = -0.7$ while varying the locality parameter.} (a) $a = 2$. (b) $a = 5$.}
\label{fig:A41}
\end{figure}

As illustrated in Fig.\ref{14a} - with $\kappa_g = -0.7$ - the effect of non-local feedback interactions is to split each Pass bands into multiple branches with opposite behavior. On Fig.\ref{2b}, for instance, each band has only one non-reciprocal branch within the corresponding BZ. Moreover, as previously reported by Rosa et al. \cite{rosa2020dynamics}, on a lattice with just one Pass bands the number of such branches is given by $a + 1$. 

Figure \ref{fig:A41} together with Fig. \ref{fig:7} depict the aforementioned phenomenon for some values of the locality parameter. %Comparing results, we 
Ii can be observed that, as $a$ increases, a branch is added to each single  PB of the diagram, confirming the above mentioned pattern and even extending it to higher frequencies. The colors denote the same bands as in Fig.\ref{fig:7} and Fig.\ref{foldedDD}.

\begin{figure}[H]
\centering
\subfigure[]{\includegraphics[width=0.495\textwidth]{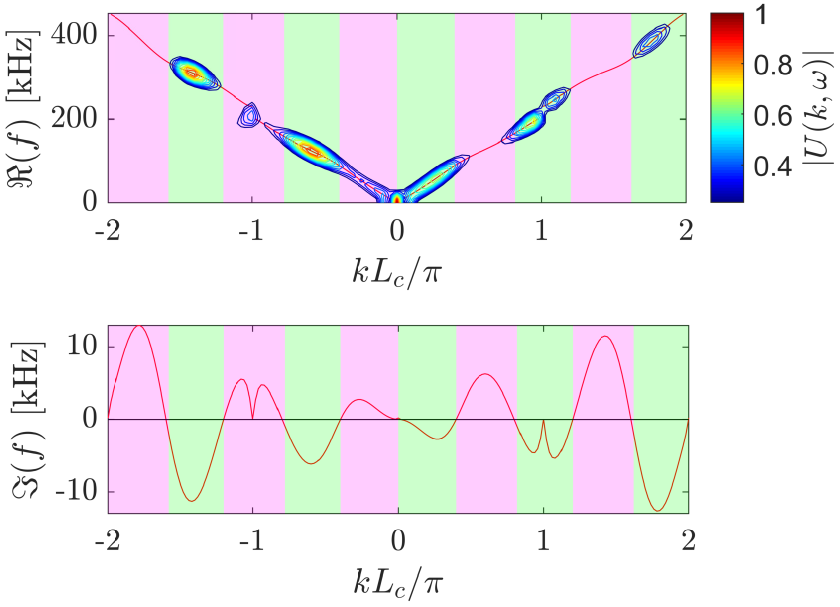}\label{15a}}
\subfigure[]{\includegraphics[width =0.495\textwidth]{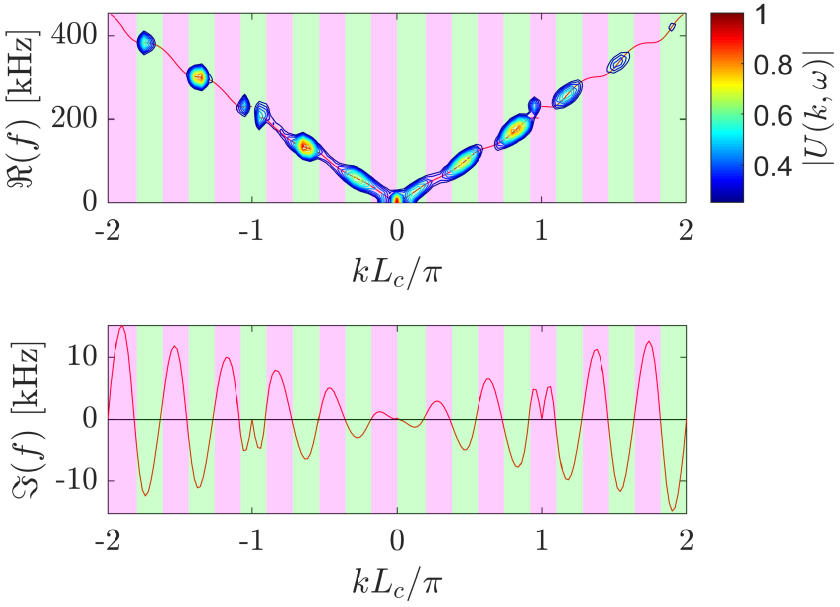}\label{15b}}
\caption{\textbf{Splitting effect of non-local feedback interactions for $\kappa_g = -0.7$ while varying the locality parameter.} (a) $a = 2$. (b) $a = 5$.}
\label{fig:A42}
\end{figure}

Figure \ref{15a} confirms the splitting effect of non-local feedback interactions, as it shows in different colors propagation (green) and stop (magenta) regions. Dispersion diagrams are truncated to show only the first two PB and curves are depicted in solid red lines. The transient response of the finite structure was simulated with the same excitation as in Fig.\ref{fig:8} and its 2DFFT contours are superposed to the diagrams of the real part and follow a colorbar normalized by the maximum element of matrix $U(k,\omega)$.

This is the version of the article before peer review or editing, as submitted by an author to Journal of Physics D: Applied Physics. IOP Publishing Ltd is not responsible for any errors or omissions in this version of the manuscript or any version derived from it. The Version of Record is available online at 10.1088/1361-6463/abf9d9.

\bibliography{ms.bib}

\bibliographystyle{ieeetr}

\end{document}